\documentclass[12pt]{article}
\usepackage{amsmath}
\usepackage{amssymb}

\input{tcilatex}

\begin{document}

\renewcommand{\thefootnote}{\fnsymbol{footnote}}

\begin{titlepage}
%---------------- preprint number & date ---------------
%\hfill\parbox{4cm}{}
%\hfill{hep-th/0407250  }
%\hfill{PUPT-2126 }\\

%\baselineskip12pt \hbox{hep-th/ }
%\baselineskip12pt \hbox{PUPT-}

\begin{flushright}
hep-th/0407250\\
PUPT-2126
\end{flushright}

%------------------------ title ------------------------
\vspace{5mm}
\baselineskip 8mm

\begin{center}
{\Large \bf The Supergravity Dual of the BMN Matrix Model}
\end{center}
\baselineskip 6mm
%---------------- authors and addresses ----------------
\vspace{10mm}
\begin{center}
Hai Lin{\footnote{hailin@princeton.edu}} \\[5mm]

{\it
Department of Physics, Princeton University\\
Princeton, NJ 08544}
\end{center}
\thispagestyle{empty}

%----------------------- abstract ----------------------
\vspace{5mm}
%\vfill

\begin{center}
{\bf Abstract}
\end{center}

We propose type IIA supergravity solutions dual to the 1/2 BPS
vacua of the BMN matrix model. These dual solutions are analyzed
using the Polchinski-Strassler method and have brane
configurations of concentric shells of D2 branes (or NS5 branes)
with various radii and D0 charge. These branes can be viewed as
polarized from $N$ D0 branes by a transverse R-R magnetic 6-form
flux and an NS-NS 3-form flux. In the region far from branes, the solutions
reduce to perturbation around the near horizon geometry of $N$ D0
branes, by turning on these R-R and NS-NS fluxes, which are dual
to the deformation of the BFSS matrix model by adding mass terms
and the Myers term. The solutions with these additional fluxes
preserve 16 supersymmetries. We also briefly discuss these fluxes
in the possible supergravity duals of M(atrix) theories on less
supersymmetric plane-waves.

\vspace{20mm}
\end{titlepage}
%-------------------------------------------------------

\baselineskip 6.5mm \renewcommand{\thefootnote}{\arabic{footnote}} %
\setcounter{footnote}{0}

%-------------------- Body of paper --------------------

\section{Introduction}

The AdS/CFT correspondence \cite{adscft} provides a remarkable
method to study the physics of gauge theories by their dual
descriptions in string theory and its low energy
limit--supergravity. In this paper, we will utilize this
correspondence to study the dual descriptions of the BMN matrix
model \cite{bmn021} and its 1/2 BPS classical vacua by IIA
supergravity solutions. This model can be considered as a 0+1
dimensional $U(N)$ gauge theory which has a discrete vacuum
spectrum and can serve as a relatively simple example of a theory
with a finite number of vacua at finite $N$. This model has very
similar vacuum structure to that of the $\mathcal{N}=1^{\ast
}$ theory \cite{n1star}, whose string theory dual was constructed
by the Polchinski-Strassler solution \cite{ps}. 

The BMN matrix model \cite{bmn021} was conjectured to be the DLCQ
of M theory on the 11-d maximally supersymmetric plane-wave background
\cite{penrose}, \cite{superposition}. The action of this model can be obtained\footnote{The action of the BMN matrix model can also be obtained by the dimensional reduction of $d=4$, $\mathcal{N}=4$ $U(N)$ SYM on ${R}\times {S}^{3}$ keeping certain $SU(2)$ invariant Kaluza-Klein modes on ${S}^{3}$ \cite{SYMS3}.}either by matrix-regularization of a supermembrane
action (e.g. \cite{supermembrane}) on the 11-d plane-wave
background, or from the quantum mechanics of $N$ D0 branes on the
background of the 11-d plane-wave compactified to 10 dimensions
\cite{BMN perturb}. The classical solutions and quantum spectrum
of this model have been extensively studied by e.g. \cite{BMN
perturb}, \cite{protect}, \cite{plefkaspectrum}. This model may
also be thought of as a deformation of the BFSS matrix model
\cite{matrixtheory} by adding mass terms and Myers term to the
Lagrangian. Due to these terms, the plane-wave background removes
the flat directions of the BFSS matrix model and the
wave-functions of the D0 branes no longer spread uniformly over
the space but are instead localized around some fuzzy spheres. The
1/2 BPS classical supersymmetric solution describes D0 branes
sitting at the origin of a 6-dimensional subspace as a result of
mass terms in these directions and in another 3-dimensional
subspace, their matrix-coordinates obey the $SU(2)$ commutation
relations as a result of the
interplay between mass terms and the Myers term \cite{bmn021}, \cite{BMN perturb}:
\begin{equation}
\lbrack X^{i},X^{j}]=i\frac{\mu }{3}\epsilon _{ijk}X^{k},\quad \quad
i,j,k=1,2,3,
\end{equation}%
where $\mu $ is a mass parameter. The coordinates $X^{i}$ ($i=1,2,3$) are $%
N\times N$ matrices and therefore their solutions are in $N$-dimensional
representations of the Lie group $SU(2)$. Since for each positive integer $n$
there is an irreducible $n$-dimensional representation of $SU(2)$, each
vacuum solution can be labeled by a partition of the integer $N$ into
positive integers $n_{i}$, with $\underset{i}{\sum }n_{i}=N$, corresponding to the direct product of these $n_{i}$-dimensional
irreducible representations. So the D0 branes form a
collection of fuzzy spheres with radii proportional to\ $n_{i}$. For large $n_{i}$, these fuzzy spheres can be well-approximated as round
spheres up to a non-commutativity correction \cite%
{kabat-taylor}.

These vacua have similar structure to those of the
$\mathcal{N}=1^{\ast }$ theory \cite{n1star}, which is a
deformation of $d=3+1,$ $\mathcal{N}=4$
$U(N)$ SYM by adding mass terms for the 3 chiral multiplets $\Phi _{i}$ (%
$i=1,2,3$) to the superpotential and the $\mathcal{N}=4$ supersymmetry is
broken to $\mathcal{N}=1$. As a result, the $F$-term equations for the
classical supersymmetric vacua yield\footnote{
The three $\Phi _{i}$ are rescaled to make the three masses equal.} \cite
{n1star}:
\begin{equation}
\lbrack \Phi _{i},\Phi _{j}]=-i\frac{m}{\sqrt{2}}\epsilon _{ijk}\Phi
_{k},\quad \quad i,j,k=1,2,3.
\end{equation}%
Since $\Phi _{i}$ are $N\times N$ traceless matrices, the solutions are also
in $N$-dimensional representations of $SU(2)$ and each vacuum is also
labeled by a partition of the integer $N$ into positive integers $n_{i}$
with $\underset{i}{\sum }n_{i}=N$, in the same way as in the BMN matrix
model.

Due to their similarities, many aspects of the BMN matrix model and the $%
\mathcal{N}=1^{\ast }$ theory may be studied in parallel both in
the field theory context and in their dual supergravity
descriptions. They can both be studied from the point of view of
deformation by relevant operators around an originally undeformed theory in the U.V. region. They both have fuzzy sphere vacua which can be interpreted
in dual supergravity solutions as branes of higher dimensionality
polarized from branes of lower dimensionality.

The various Polchinski-Strassler solutions \cite{ps}
in the string dual of the $\mathcal{N}%
=1^{\ast }$ theory have brane configurations corresponding to D3
branes polarized into various D5 branes (in the weak coupling
regime) or NS5
branes (in the strong coupling regime) via Myers' dielectric effect \cite%
{mye053}. On the near horizon geometry of $N$ D3 branes, i.e.
$AdS_{5}\times S^{5}$, Polchinski and Strassler \cite{ps} found
that one can turn on additional R-R 3-form fluxes and NS-NS 3-form fluxes to
polarize the D3 branes into D5 or NS5 branes with world-volumes ${R}%
^{4}\times {S}^{2}$. They found that the radii of the
${S}^{2}$ of these D5 branes are proportional to the
D3-charge that these D5 branes carry, under certain
approximations. 

In this paper, we find similar physics happened in terms of D0
branes. We start from the near horizon geometry of $N$ D0 branes
which is dual to the BFSS matrix model \cite{D0dual}, and find
that one can turn on additional transverse magnetic R-R 6-form
flux and NS-NS 3-form flux whose Hodge duals can couple to D2 or
NS5 charge and thereby to cause the $N$ D0 branes to polarize into
various D2 or NS5 branes. We find that the D2 branes polarized
from the
D0 branes in the presence of these additional fluxes have ${S}^{2}$%
-wrapped world-volumes and the radii of their $%
{S}^{2}$ are proportional to the D0-charge they carry,
under certain approximations. We find that the general equilibrium
brane configuration could consist of many concentric D2 branes
each with its radius proportional to its D0-charge.

We thus propose a holographic dual description of the 1/2 BPS classical vacua of the BMN matrix model, using dual IIA solutions with brane configurations, in the cases when all $n_i$ are large. In
the appropriate regimes of parameters, there is a one-to-one
correspondence between the supergravity solutions in the bulk,
where there are concentric branes carrying D0-charge, and the 1/2
BPS classical vacua of the BMN matrix model on the boundary, which
are collections of fuzzy spheres. The concentric branes are either
D2 branes or NS5 branes, in the regimes of weak or strong
effective 't Hooft couplings, respectively, in the matrix
perturbation theory of the BMN matrix model \cite{BMN perturb}. In
the D2 brane descriptions, each dual supergravity solution
corresponds to a way of dividing up and distributing the total
D0-charge $N$ to several D2 branes each with D0-charge $n_{i}$, by
an identical partition of $N$ in the matrix model side, in terms
of fuzzy spheres. On the other hand, in the NS5 brane
descriptions, each dual supergravity solution also corresponds to
a way of dividing up and distributing the total D0-charge $N$ to
several NS5
branes, but by a dual partition\footnote{
Here, a dual partition of $N$ is defined via switching the rows
and columns of a Young tableau, see e.g. \cite{transverse5} p.6.}
of $N$ in the matrix model side. These concentric branes in the
supergravity side are holographically mapped to the fuzzy spheres
in the matrix model side. In the
large $r$ region of the supergravity solutions (where $r$ is the
radial variable of the 9 spatial dimensions), the additional R-R
6-form flux and NS-NS 3-form flux are dual to the deformation of
the BFSS matrix model by adding mass terms and the Myers term.

These solutions of IIA when lifted up to 11 dimensions describe
supergravitons polarized into M2 or M5 branes. They are giant gravitons
(e.g. \cite{INVASION}, \cite{determinant}, \cite{CJR}) each carrying a
fraction of the total light-cone momentum. The light-cone momentum of the $N$
supergravitons $p_{+}=\frac{N}{R}$ are shared to several M2 branes each with
light-cone momentum $p_{+}^{(i)}=\frac{n_{i}}{R}$ and with radius
proportional to $n_{i}$, in the same way as a partition of integer $N$. In
the M5 brane description, it is also a way of sharing the total light-cone
momentum to several M5 branes but by a dual partition of integer $N$ \cite%
{transverse5}.

The main body of the paper will focus on the details of construction of the dual supergravity solutions in terms of D2
brane configurations by the Polchinski-Strassler method, valid in the regimes of weak effective 't Hooft couplings in the matrix perturbation theory of the BMN
matrix model \cite{BMN perturb}. In the next section, we study the R-R 6-form flux $G_{6}$ and NS-NS 3-form flux $H_{3}$, in the large $r$ region, as
perturbation around the near horizon geometry of $N$ D0 branes.
In section 3, we study the equilibrium radii of the ${S}^{2}$
-wrapped D2 branes with D0-charge in the solutions with general
brane configurations. In section 4, we solve the fluxes $H_{3}$ and $G_{6}$
as being sourced by these polarized D0 branes with D2-charge. In section 5, we study the metric and dilaton near each shell of
the branes as well as in the large $r$ region. In the last section, we discuss related issues to our results and also possible generalizations to the supergravity duals of M(atrix) theories on less-supersymmetric plane-waves.

\bigskip

\section{The R-R 6-form flux and NS-NS 3-form flux in the large $r$ region}

Since the BMN matrix model can be considered as a deformation of
the BFSS matrix model, its dual supergravity solutions, in the
large $r$ region, can be considered as perturbation around the near
horizon geometry of $N$ D0 branes, which in string frame is
(e.g. \cite{pbrane}, \cite{duff}):
\begin{eqnarray}
ds^{2} &=&-Z^{-1/2}dt^{2}+Z^{1/2}d\overrightarrow{x}_{i}^{2},\quad \quad
\quad i=1,...,9,  \notag \\
e^{\Phi } &=&g_{s}Z^{3/4},\quad \quad C_{1}=g_{s}^{-1}(Z^{-1}-1)dt,\quad
\quad Z=\frac{R^{7}}{r^{7}},  \label{d0}
\end{eqnarray}%
where $R^{7}=60\pi ^{3}g_{s}N\alpha ^{^{\prime }7/2}$.

The fluctuations around this background we are interested in are the R-R flux $\widetilde{F}_{4}$ and NS-NS flux $H_{3}$ which are relevant to the couplings to the D2 or NS5 branes that can be polarized from D0 branes.
When these additional fluxes in large $r$ region can be considered as small
fluctuations around the above background (\ref{d0}), it's easy to see that
the perturbations of the metric, dilaton and $F_{2}$ are all of
second order or higher in the fluctuations. Therefore if we neglect quantities that are of second order or higher in the fluctuations, we only need to turn on
these additional fluxes without giving corrections to the background. For convenience, we can dualize the 4-form flux $\widetilde{F}_{4}$ into a transverse\footnote{By transverse we mean that the forms such as $H_{3} $ and $G_{6}$ etc. have all components transverse to the D0 brane world-volume.} 6-form flux $G_{6}$ via $G_{6}=Z^{3/8}\ast
\widetilde{F}_{4}$, where $\ast$ is the Hodge dual with respect to the 10d metric in Einstein frame. After some derivation in appendix A, the equations of motion for the transverse fluxes $H_{3}$ and $
G_{6}$ turned out to possess a simple form:
\begin{eqnarray}
dH_{3} &=&0,  \label{h3close} \\
dG_{6} &=&0,  \label{g6close}
\end{eqnarray}%
\begin{eqnarray}
d[Z^{-1}(H_{3}-g_{s}\ast _{9}G_{6})] &=&0,  \label{harmonic3} \\
d[Z^{-1}(\ast _{9}H_{3}-g_{s}G_{6})] &=&0,  \label{harmonic6}
\end{eqnarray}%
where $\ast _{9}$ is the Hodge dual in the transverse 9-d with respect to a flat 9-d metric. These constraints tell us that $H_{3}$ and $G_{6}$ are both
closed forms and $Z^{-1}(H_{3}-g_{s}\ast _{9}G_{6})$ is annihilated by both $%
d$ and $d\ast _{9}$ in the transverse 9 dimensions. 

The solution should break the isometry $SO(9)$ to $SO(3)\times SO(6)$, where
the $SO(3)$ is the isometry of the 123 subspace and the $SO(6)$ is
the isometry of the other 6-d transverse subspace. According to
this isometry, in the large $r$, we should look for the fluxes of
the form 
\begin{eqnarray}
H_{3} &=&r^{m}(\alpha T_{3}+\beta V_{3}),  \label{flux-ansat-h3} \\
G_{6} &=&r^{n}(\gamma \ast _{9}T_{3}+\delta \ast _{9}V_{3}),
\label{flux-ansat-g6}
\end{eqnarray}%
where $T_{3}=dx^{1}\wedge dx^{2}\wedge dx^{3}$ is the volume form of the 123
subspace and $V_{3}$ is defined as $V_{3}=d\ln r\wedge S_{2}$, where $S_{2}=%
\frac{1}{2!}\varepsilon _{ijk}x^{i}\wedge dx^{j}\wedge dx^{k}$, ($%
i,j,k=1,2,3 $) and $m,n,\alpha ,\beta ,\gamma ,\delta $ are constants. 

By plugging this ansatz, the set of equations (\ref{h3close})-(\ref{harmonic6})
admit four solutions in two pairs (see appendix B). Each pair consists of one
non-normalizable and one normalizable solution as follows\footnote{
The solution with the $n=0$ fluxes in (\ref{n=0}), when uplifted
to 11d, leads to the solution of the \textquotedblleft
superposition" of the 11d gravitational wave
and 11d plane-wave, as described in \cite{superposition} p.20, \cite%
{superp}.}:
\begin{eqnarray}
\text{\textrm{first} }\mathrm{pair} &\mathrm{:}&\text{ }  \notag \\
H_{3} &=&\alpha r^{-7}(T_{3}-\frac{7}{3}V_{3}),\quad \quad
G_{6}=g_{s}^{-1}\alpha r^{-7}(\frac{1}{3}\ast _{9}T_{3}-\frac{7}{3}\ast
_{9}V_{3}),  \label{n=-7} \\
H_{3} &=&\alpha r^{-9}\left( T_{3}-3V_{3}\right) ,\quad \quad
G_{6}=g_{s}^{-1}\alpha r^{-9}(\ast _{9}T_{3}-3\ast _{9}V_{3}),  \label{n=-9}
\end{eqnarray}%
\begin{eqnarray}
\ \quad \text{\textrm{second} }\mathrm{pair} &\mathrm{:}&\   \notag \\
H_{3} &=&\alpha T_{3},\quad \quad \quad \ \ \ \ \ \ \ \ \ \ \ \ \ \
G_{6}=g_{s}^{-1}\alpha \ast _{9}T_{3},  \label{n=0} \\
H_{3} &=&\alpha r^{-16}(T_{3}-\frac{16}{3}V_{3}),\quad
G_{6}=g_{s}^{-1}\alpha r^{-16}(-\frac{5}{3}\ast _{9}T_{3}+\frac{8}{3}\ast
_{9}V_{3}),  \label{n=-12}
\end{eqnarray}%
where the $\alpha $ in different lines are different. In the language of AdS/CFT correspondence, the pair of
$n=-7$ and $n=-9$ solutions\footnote{Here $n$ is the scaling dependence of the fluxes
on $r$ in large $r$ region.}, i.e. (\ref{n=-7}), (\ref{n=-9}),
corresponds to the operators of mass
deformation in the matrix model side and the VEV of it respectively \cite%
{boundarybulk}. As will be shown in section 4, in the
large $r$ region, $H_{3}$
and $G_{6}$ are the superpositions of the $n=-7$ and $n=-9$ solutions, while the $%
n=-7$ solution is of first order in the mass parameter $\mu $ of
the BMN matrix model, and the $n=-9$ solution is of third order in
the mass parameter $\mu $ (see appendix F).

Since our solutions are dual to 1/2 BPS classical vacua of BMN
matrix model, we should have 16 supersymmetries in our solutions.
This is one of the differences between the dual solutions of fuzzy
sphere vacua of BMN matrix model and those of the
$\mathcal{N}=1^{\ast }$ theory. In the Polchinski-Strassler case,
the supersymmetry is broken from $\mathcal{N}=4$ to
$\mathcal{N}=1$ \cite{grana}, \cite{pssusynew}, while in our case
the solutions after turning on $H_{3}$ and $G_{6}$ still preserve
16 supersymmetries.

We therefore explicitly solved the Killing spinor perturbatively in first
order in $\mu $, when we turn on the fluctuations of the $n=-7$ solution (\ref%
{n=-7}) of $H_{3}$ and $G_{6}$, which are of first order in $\mu
$. The Killing spinor before perturbation is the Killing spinor
$\epsilon ^{(0)}$ in the near horizon geometry of $N$ D0 branes,
and the Killing spinor, after turning on $H_{3}$ and $G_{6}$ that
are of first order in $\mu $, could be written perturbatively as
$\epsilon =\epsilon ^{(0)}+\epsilon ^{(1)}$, where $\epsilon
^{(1)}$ is the perturbation of the Killing spinor, and is of first
order in $\mu $.

We can thereby split the gravitino equations and the dilatino
equations order by order in $\mu $, and the equations for the
first two orders in $\mu $ are\footnote{The letters with a slash denote the contractions of forms with gamma matrices, for example: ${\not{F}}_{2}=\frac{1}{2!}F_{ab}\Gamma ^{ab},{\not{H}}_{3}=\frac{1}{3!}
H_{abc}\Gamma ^{abc},\widetilde{\not{F}}_{4}=\frac{1}{4!}\widetilde{F}
_{abcd}\Gamma ^{abcd}$.} (in string frame \cite{susyrule},
see appendix C):
\begin{eqnarray}
\left( \frac{1}{2}\Gamma ^{m}\partial _{m}\Phi +\frac{3}{8}e^{\Phi }\not{F%
}_{2}\Gamma ^{\underline{11}}\right) \epsilon ^{(0)} &=&0,
\label{dilatino-zero-order} \\
\left( \partial _{m}+\frac{1}{4}\omega _{mab}\Gamma ^{ab}+\frac{1}{8}e^{\Phi
}\not{F}_{2}\Gamma _{m}\Gamma ^{\underline{11}}\right) \ \epsilon ^{(0)}
&=&0,  \label{gravitino-zero-order}
\end{eqnarray}%
\begin{eqnarray}
\left( \frac{1}{2}\Gamma ^{m}\partial _{m}\Phi +\frac{3}{8}e^{\Phi }\not{F%
}_{2}\Gamma ^{\underline{11}}\right) \epsilon ^{(1)} &=&-\left( \frac{1}{4}%
{\not H}_{3}\Gamma ^{\underline{11}}+\frac{1}{8}e^{\Phi }\widetilde{%
\not{F}}_{4}\right) \epsilon ^{(0)},  \label{dilatino-1st-order} \\
\left( \partial _{m}+\frac{1}{4}\omega _{mab}\Gamma ^{ab}+\frac{1}{8}e^{\Phi
}\not{F}_{2}\Gamma _{m}\Gamma ^{\underline{11}}\right) \epsilon ^{(1)}
&=&-\left( \frac{1}{8}H_{mab}\Gamma ^{ab}\Gamma ^{\underline{11}}+\frac{1}{8}%
e^{\Phi }\widetilde{\not{F}}_{4}\Gamma _{m}\right) \epsilon ^{(0)}.
\label{gravitino-1st-order}
\end{eqnarray}%

The first pair of equations (\ref{dilatino-zero-order}), (\ref%
{gravitino-zero-order}), i.e. the zeroth order equations, give the
unperturbed $\epsilon ^{(0)}$ which is known in the literature (e.g. \cite{g=1}, \cite%
{d0multipole}):
\begin{equation}
\epsilon ^{(0)}=Z^{-1/8}\eta ,
\end{equation}%
where $\eta$ is a constant spinor with definite helicity via
projection condition $(1+\Gamma ^{\underline{0}}\Gamma
^{\underline{11}})\eta =\eta $.
Input this result into the other pair of equations (\ref{dilatino-1st-order}%
), (\ref{gravitino-1st-order}), i.e. the first order equations, we find that
the first order correction $\epsilon ^{(1)}$ can be separated into a
time-dependent part $\epsilon _{1}^{(1)}$ and a time-independent part $%
\epsilon _{2}^{(1)}$, and the former has the same helicity with the
unperturbed $\epsilon ^{(0)}$ while the latter has the opposite helicity to $%
\epsilon ^{(0)}$. 

Explicit calculations are in appendix C and the results
are\footnote{In the expressions of the Killing spinors (\ref{eps1result}), (\ref
{eps2result}), the various indices are: The indices $l,m,n$ denote
1,...,9; the indices $i$ denotes 1,2,3 and the indices $a$ denotes
4,...,9; the indices of gamma matrices with a bar below are the
gamma matrices in tangent space. Here we consistently use another
parameter $\widetilde{\mu }$ in all the expressions instead of the
mass parameter $\mu $ in BMN matrix model. They are proportional
to each other, i.e. $\widetilde{\mu }=\zeta \mu $, where $\zeta $
is a dimensionless factor of order 1, which may be figured out by
comparing the D2 potential (in section 3) with the matrix model
action.}
\begin{equation}
\epsilon =\epsilon ^{(0)}+\epsilon _{1}^{(1)}+\epsilon _{2}^{(1)},
\end{equation}%
\begin{equation}
\epsilon _{1}^{(1)}=\frac{1}{12}\cdot \frac{1}{3!}Z^{-1}[H_{lmn}-g_{s}(\ast
_{9}G_{6})_{lmn}]\Gamma ^{\underline{l}\underline{m}\underline{n}}\epsilon
^{(0)}t=-\frac{1}{12}\widetilde{\mu }(\Gamma ^{\underline{1}\underline{2}%
\underline{3}}t)\epsilon ^{(0)},  \label{eps1result}
\end{equation}%
\begin{equation}
\epsilon _{2}^{(1)}=\widetilde{\mu }(\frac{1}{6}\Gamma ^{\underline{i}}x^{i}-%
\frac{1}{12}\Gamma ^{\underline{a}}x^{a})Z^{1/2}\Gamma ^{\underline{1}%
\underline{2}\underline{3}}\Gamma ^{\underline{0}}\epsilon ^{(0)},
\label{eps2result}
\end{equation}%
where the fluxes in first order in $\mu $ are
\begin{equation}
H_{3}=\frac{3}{2}\widetilde{\mu }Z(-T_{3}+\frac{7}{3}V_{3}),\quad \quad
G_{6}=\frac{3}{2}g_{s}^{-1}\widetilde{\mu }Z(-\frac{1}{3}\ast _{9}T_{3}+%
\frac{7}{3}\ast _{9}V_{3}),  \label{n=-7flux}
\end{equation}%
and $Z=\frac{R^{7}}{r^{7}}$. 

We see from (\ref{n=-7flux}) that the leading terms of the fluxes $H_{3}$ and $%
G_{6}$ in the large $r$ region are proportional to the total
D0-charge $N$ of the branes and are independent of the specific
brane configurations in the small $r$ region. Note that the
equation for the time-dependent part $\epsilon _{1}^{(1)}$ yields
a constraint that the components $Z^{-1}[H_{lmn}-g_{s}(\ast
_{9}G_{6})_{lmn}]$ should be constants and therefore
$Z^{-1}(H_{3}-g_{s}\ast _{9}G_{6})$ need to be a constant 3-form
in the transverse 9-d (appendix C). This is indeed the case, since
it equals to $-\widetilde{\mu }T_{3}$.

So we have checked that our solutions with the fluxes
in large $r$ region in first order in $%
\mu $, i.e. the $n=-7$ solution, are supersymmetric, preserving
the 16 supersymmetries. In section 4, we will see that the
$n=-9$ solution is of third order in $\mu $ in large $r$ region.
In order to check the supersymmetry when superposing the $n=-7$
and $n=-9$ in large $r$ region, we
need to give second and third order corrections to the metric, dialton and $%
F_{2}$. One difference from the original Polchinski-Strassler
solutions \cite{ps} is that we still preserve all the supersymmetries of the
unperturbed background after adding on the additional fluxes. This
is not surprising since the $T_{3}$ in the expressions of our
fluxes is maximally symmetric under
the isometry $SO(3)\times SO(6)$. The expressions for the fluxes $H_{3}$ and $%
G_{6},$ as well as the Killing spinor in this section, are valid
in large $r$ region where the additional fluxes can be considered
as small fluctuations compared to the background. The $H_{3}$ and
$G_{6}$ near the brane sources will be discussed in section 4.

\bigskip

\section{The position of D2 branes with D0 charge in the small $r$
region}

In last section, we have studied the fluxes in the large $r$
region, which are dual to the operators of deformation in matrix
model side. In this section, we will consider the small $r$
region, where there are branes which are the holographic maps of
the fuzzy spheres in the matrix model side. The
general brane configurations in our solutions are concentric shells of $p_{i} $ D2 branes each with $q_{i}$ D0-charge, where $i$ denotes
the $i$th shell, and $\underset{i}{\sum }p_{i}q_{i}=$ $N$. We will show that the radii of these D2 branes in our solution are proportional to the D0-charge $q_{i}$ that they carry, under certain approximation.

In a general brane configuration corresponding to the partition $N=\underset{i}{\sum }p_{i}q_{i}$, each brane is in an equilibrium
position under the potential it feels in the presence of all
branes. The equilibrium radius of each brane corresponds to the
location of the minimum of the potential that the brane feels. 
Before we calculate the potential that each brane feels in a
general brane configuration, we will first solve
a simpler problem. We will calculate the potential of a probe D2
brane with D0-charge $q$ in the background of the near horizon
geometry of single-center $N$ D0 branes, with the $n=-7$
additional fluxes $H_{3}$ and $G_{6}$, i.e. (\ref{n=-7flux}),
turned on. Then we can generalize the result for the probe brane
to the branes that are not probes, in a general brane
configuration.

The reason we can make this generalization is that there is
certain configuration-independence in our solutions, for example,
as showed in section 2, the 3-form $Z^{-1}(H_{3}-g_{s}\ast
_{9}G_{6})$ remains a constant form
at large $r$, independent of what the brane configurations are in the small $%
r$ region. In this section, we will show, under certain
approximations, i.e. condition (\ref{chargecondition}), the brane
potential is also
configuration-independent. It only depends on the radius of the ${S}%
^{2}$ of the brane and the D0-charge of it, and doesn't care about
how the other branes distribute, under this approximation.

In the probe calculation, we require that the D0-charge $q$ it
carries is much smaller than the background D0-charge $N$ so that
it can be treated as a probe. However, we can relax this condition
later when we make the generalization, due to the
configuration-independence. The brane will take the shape of an
${S}^{2}$ embedded in 123 subspace and at the origin of the
other six transverse dimensions. The DBI and WZ action of the D2
brane with $q$ units of D0-charge in the string frame is
\begin{equation}
S_{D2}=-\tau _{D2}\int d^{3}\sigma e^{-\Phi }\sqrt{-\det (G_{\alpha \beta
}+2\pi \alpha ^{^{\prime }}\mathcal{F}_{\alpha \beta })}+\tau _{D2}\int
(C_{3}+2\pi \alpha ^{^{\prime }}\mathcal{F}_{2}\wedge C_{1}),
\label{d2action}
\end{equation}%
where $2\pi \alpha ^{^{\prime }}\mathcal{F}_{2}=2\pi \alpha ^{^{\prime
}}F_{2}-B_{2}$. We choose the static gauge that the world-volume coordinates
are the same as the space-time ones, i.e. $t,\theta ,\varphi ,$ where the
angles parameterize the ${S}^{2}$. The radius of the ${S}^{2}$
in 123 subspace is denoted as $r_{1}$. The D0-charge $q$ of the D2 brane is
realized as a world-volume 2-form flux $F_{2}=\frac{1}{2}q\sin \theta
d\theta \wedge d\varphi $. The $B_{2}$ is given by the $n=-7$ solution (\ref%
{n=-7flux})\ of the $H_3$ in section 2. 

We will make an
approximation that is similar to Polchinski-Strassler \cite{ps}
that the dominating terms in both DBI and WZ actions are from the
contribution of $F_{2}$, which requires the conditions (see appendix D):
\begin{equation}
\frac{4\pi ^{2}\alpha ^{^{\prime }2}\det F_{2}}{\det G_{\perp }}=\frac{\pi
^{2}\alpha ^{^{\prime }2}q^{2}r_{1}^{3}}{R^{7}}\sim \frac{q^{2}r_{1}^{3}}{%
g_{s}N}\gg 1,\quad \ \ \ \ \ \frac{2\pi \alpha ^{^{\prime }}F_{\alpha \beta }%
}{B_{\alpha \beta }}=-\frac{2\pi \alpha ^{^{\prime }}qr_{1}^{4}}{\widetilde{%
\mu }R^{7}}\sim \frac{qr_{1}^{4}}{g_{s}N}\gg 1,  \label{condition1}
\end{equation}%
where $G_{\perp }$ is the pull-back metric parallel to the ${S}^{2}$.
Under this approximation we can expand the square-root in the DBI action
around $\det F_{2}$ and then the leading terms in the DBI and WZ actions
precisely cancel (appendix D):
\begin{equation}
-\tau _{D2}g_{s}^{-1}\int d^{3}\sigma Z^{-3/4}\sqrt{-\det G_{\shortparallel }%
}\cdot 2\pi \alpha ^{^{\prime }}\sqrt{\det F_{2}}+\tau _{D2}\int 2\pi \alpha
^{^{\prime }}F_{2}\wedge C_{1}=0,
\end{equation}%
where we choose the gauge $C_{1}=g_{s}^{-1}Z^{-1}dx^{0}$ and $%
G_{\shortparallel }$ is the pull-back metric parallel to time.
This is because that the leading terms in the DBI and WZ actions,
which are both contributed from the $F_{2}$, describe the
potential between D0 and D0 charges and it should be zero due to
supersymmetry. The two leading terms are both large but they
cancel.

The subleading terms of the DBI and WZ actions are respectively:
\begin{equation}
-\tau _{D2}g_{s}^{-1}\int d^{3}\sigma Z^{-3/4}\frac{\sqrt{-\det
G_{\shortparallel }}\det G_{\perp }}{4\pi \alpha ^{^{\prime }}\sqrt{\det
F_{2}}}=\int dt\left( -\frac{2\tau _{D2}r_{1}^{4}}{g_{s}\alpha ^{^{\prime }}q%
}\right) ,
\end{equation}%
\begin{equation}
\tau _{D2}\int C_{3}=\int dt\left( \frac{4\pi \tau _{D2}\widetilde{\mu }%
r_{1}^{3}}{3g_{s}}\right) ,
\end{equation}%
where $C_{3}=\frac{1}{3}g_{s}^{-1}\widetilde{\mu }dx^{0}\wedge S_{2}$.\
These two terms in the brackets are of order $r_{1}^{4}$ and $\widetilde{\mu
}r_{1}^{3}$ respectively, which can be identified with the commutator term $%
\mathrm{Tr}\left( \frac{1}{4}[X^{i},X^{j}]^{2}\right) $ and the Myers term $-%
\mathrm{Tr}\left( i\frac{\mu }{3}\epsilon _{ijk}X^{i}X^{j}X^{k}\right) $ in
the matrix model Lagrangian. There should be another term of order $%
\widetilde{\mu }^{2}r_{1}^{2}$ coming from the second order corrections of
the metric, dilation and $C_{1}$, which can be identified as $-\mathrm{Tr}$ $%
\frac{1}{2}\left( \frac{\mu }{3}\right) ^{2}\left( X^{i}\right)
^{2}$ in the matrix model Lagrangian. The subleading terms of the
DBI and WZ actions, plus the second order corrections from the
metric, dilation and $C_{1}$ are expected to complete the
potential with a perfect square, due to supersymmetry condition.
This is not surprising since this corresponds to
the perfect-square term $-\int dt$ $\frac{1}{2}\mathrm{Tr}\left( \frac{\mu }{%
3}X^{i}+i\epsilon ^{ijk}X^{j}X^{k}\right) ^{2}$ in the matrix model action
\cite{BMN perturb}. So the third term could be in principle read off from
the first two terms. 

By the approximation (\ref{condition1})\ and
supersymmetry condition, the action should then be
\begin{equation}
S\approx -\int dt\frac{2\tau _{D2}}{g_{s}\alpha ^{^{\prime }}q}\left(
r_{1}^{2}-\frac{\pi }{3}\widetilde{\mu }\alpha ^{^{\prime }}qr_{1}\right)
^{2}.  \label{braneaction}
\end{equation}%
So we see that the brane potential depends on the radius of the ${S}^{2},$ i.e. $%
r_{1}$, and its D0-charge $q$, and is independent of the
warp-factor $Z$ under the approximation (\ref{condition1}). The
subleading term in DBI action is $Z$-independent since both $Z$
factors from numerator and denominator exactly cancel (appendix
D). The subleading term in WZ action is also $Z$-independent
because $C_{3}$ is $Z$-independent. And the third term should also
be $Z$-independent due to supersymmetry. Therefore the potential
of the brane only cares about its D0-charge and is independent of
the brane configuration of the solution under this approximation.
The potential of a non-probe D2 brane with D0-charge in a general
brane configuration is thus still of the form (\ref{braneaction}).
There is a non-trivial equilibrium radius at
\begin{equation}
r_{0}\approx \frac{\pi }{3}\widetilde{\mu }\alpha ^{^{\prime }}q,
\end{equation}%
which is proportional to the D0-charge $q$. This is similar to the matrix
model description of the fuzzy spheres that each of them has a radius proportional to the size $q$ of its block matrix-coordinates \cite{bmn021}, \cite{BMN perturb}, up to a non-commutativity correction \cite{kabat-taylor}.

After knowing the equilibrium radius $r_{0}$,
we get the consistent condition for our approximation from (%
\ref{condition1}) that
\begin{equation}
\frac{q^{5}}{g_{s}N}\gg 1.  \label{chargecondition}
\end{equation}%
This is a kind of scaling bound between $q$ and $g_{s}N$, where
$q$ is the D0-charge of the D2 brane and $N$ is the background
D0-charge\footnote{This is very similar to the condition for the
approximation in the original Polchinski-Strassler solution, i.e.
$\frac{n^{2}}{g_{s}N}\gg 1$ \cite{ps}, where $n$ was the D3-charge
of the D5 brane and $N$ was the background D3-charge. We have
different powers of $n$ (or $q$) because the powers of the $r$ in
the warp-factor $Z$ is different. In their case $Z=\frac{R^{4}}{r^{4}}$, while in our case $Z=\frac{R^{7}}{r^{7}}$. The warp-factor dilutes the background charge so the power dependence of $n$ (or $q$) for different $Z$ is
different.}. 
The approximation (\ref{chargecondition}) actually
guarantees that the contributions to the DBI and WZ actions are
both from the D0-charge but they cancel and then the remaining
terms form a perfect square and is independent of the brane
configuration and the warp-factor.

So in our general brane configurations, suppose there are concentric shells
of $p_{i}$ D2 branes each with D0-charge $q_{i}$ (the label $i$ denotes the $%
i$th shell), the total potential of all the branes can be considered as the
sum of individual potentials and we have:
\begin{equation}
S\approx -\int dt\underset{i}{\sum }\frac{2\tau _{D2}p_{i}}{g_{s}\alpha
^{^{\prime }}q_{i}}\left( \left( r_{1}^{(i)}\right) ^{2}-\frac{\pi }{3}%
\widetilde{\mu }\alpha ^{^{\prime }}q_{i}r_{1}^{(i)}\right) ^{2},
\end{equation}%
where $r_{1}^{(i)}$ denote the radii of the ${S}^{2}$ of the branes
on the $i$th shell and their equilibrium radii $r_{0}^{(i)}$ are therefore%
\begin{equation}
r_{0}^{(i)}\approx \frac{\pi }{3}\widetilde{\mu }\alpha ^{^{\prime
}}q_{i},\quad \quad \quad \quad \quad \quad \underset{i}{\sum }p_{i}q_{i}=N.
\label{radius}
\end{equation}%
Note that there could be coincident $p_{i}$ D2 branes on the same
shell if they have the same amount of D0-charge $q_{i}$, which
corresponds to some copies of the irreducible representations of
the $SU(2)$ of the same dimension, in the matrix model side.

\bigskip

\section{The additional fluxes in the presence of polarized sources}

In section 2, we focused on the large $r$ region and found out the
additional fluxes $H_{3}$ and $G_{6}$, as perturbations around the
near horizon geometry of $N$ D0 branes. In section 3, we have
studied the situations in the small $r$ region and figured out the
radius of each D2 brane, in general configurations. The leading terms of these
fluxes $H_{3}$ and $G_{6}$ in the large $r$ region depend on the total D0-charge of all the branes and are independent of the brane configurations,
while in the small $r$ region these fluxes are dependent on specific brane
configurations. The expressions for the fluxes in section 2 are
not valid near the brane sources. In this section we will
therefore study the behavior of $H_{3} $ and $G_{6}$ in the
presence of these sources.

For simplicity of the calculation, we study the special case when
there is only a single shell of D2 branes with total D0-charge $N$. Suppose we have $p$ coincident D2 branes each with D0-charge $q=N/p$, so the
radius of the shell is $r_{0}\approx \frac{\pi }{3}\widetilde{\mu
}\alpha ^{^{\prime }}(N/p).$ We study the case that the total D2-charge
$p$ is small, so the background metric can be approximated by the
near horizon geometry of multi-center D0 branes distributed on the
$S^2$ with radius $r_0$.

The equation for $\ast \widetilde{F}_{4}$ will have a source term
due to the D2-charge. We are interested in the $H_{3}$ and $G_{6}$
on the background of the near horizon geometry of a shell of
multi-center D0 branes with small D2-charge turned on. We are now doing perturbation in terms of small parameter $p$. The
warp-factor $Z=\frac{R^{7}}{r^{7}}$ is replaced by the
multi-center warp factor $Z_{1}$ in solution (\ref{d0}):%
\begin{equation}
Z_{1}=\frac{R^{7}}{10r_{1}r_{0}}\left[ \frac{1}{%
[(r_{1}-r_{0})^{2}+r_{2}^{2}]^{5/2}}-\frac{1}{%
[(r_{1}+r_{0})^{2}+r_{2}^{2}]^{5/2}}\right] ,  \label{z1oneshell}
\end{equation}%
which is the superposition of harmonic functions, corresponding to
D0-charge uniformed distributed on an ${S}^{2}$ with radius
$r_{0}$ in the 123 subspace and centered at the origin of the
other 6-d transverse subspace. Here $r_{1}$ is the radius of 123
subspace and $r_{2}$ is the radius of the other 6-d transverse
subspace. Its easy to see that $Z_{1}$ reduces to $Z=\frac{R^{7}}{r^{7}}$ at large $r$.

The IIA SUGRA equations for $H_{3}$ and $G_{6}$ on the background
of the near horizon geometry of a single shell of multi-center D0
branes with radius $r_0$ and small D2-charge $p$ should be (see appendix E):
\begin{eqnarray}
dH_{3} &=&0,  \label{h3source} \\
dG_{6} &=&J_{7},  \label{g6source}
\end{eqnarray}%
\begin{eqnarray}
d[Z_{1}^{-1}(H_{3}-g_{s}\ast _{9}G_{6})] &=&0,  \label{harmonic3multi} \\
d[Z_{1}^{-1}(\ast _{9}H_{3}-g_{s}G_{6})] &=&0,  \label{harmonic6multi}
\end{eqnarray}%
where there is a source term $J_7$ due to D2-charge:
\begin{equation}
J_{7}=2\kappa ^{2}\tau _{D2}g_{s}^{-2}p\delta (r_{1}-r_{0})\delta
^{6}(r_{2})dr_{1}\wedge dr_{2}\wedge \omega _{5},
\end{equation}%
where $\delta ^{6}(r_{2})=\delta (x_{4})\delta (x_{5})\delta (x_{6})\delta
(x_{7})\delta (x_{8})\delta (x_{9})$ and $\omega _{5}=r_{2}^{5}\cdot \mathrm{%
dvol}({S}^{5})$, where $\mathrm{dvol}({S}^{5})$ denotes the
volume-form of an ${S}^{5}$ with unit radius embedded in 456789
subspace and centered at origin. The source term
appears in the Bianchi identity but not in the equations for $
Z_{1}^{-1}(H_{3}-g_{s}\ast _{9}G_{6})$ and its 9-d Hodge dual\footnote{Since in our case the forms $H_{3}$ and $F_{4}$ both have overall factors of the volume-form of the ${S}^{2}$
in 123 subspace due to the isometry $SO(3)\times SO(6)$, the terms $%
F_{4}\wedge H_{3}$ and $F_{4}\wedge F_{4}$ are zero and drop off on the
right sides of (\ref{g6source}), (\ref{harmonic6multi}).}.

The equations for $Z_{1}^{-1}(H_{3}-g_{s}\ast _{9}G_{6})$\ remain the same as in (\ref%
{harmonic3}), (\ref{harmonic6}), just with $Z$ replaced by $Z_1$. Furthermore, since when $r$ goes to infinity, this
form is a constant form, we infer that the harmonic form $%
Z_{1}^{-1}(H_{3}-g_{s}\ast _{9}G_{6})=-\widetilde{\mu }T_{3}$.
Although $H_{3},G_{6},Z_{1}$ all depend on the brane
configurations, the combination $Z_{1}^{-1}(H_{3}-g_{s}\ast
_{9}G_{6})$ is independent of the brane configurations. This also
results in that the potential $C_{3}$ is independent of the
warp-factor $Z_{1}$ and the
brane configurations. The difference between equations (\ref{h3close})-(\ref%
{harmonic6}) and (\ref{h3source})-(\ref{harmonic6multi}), besides
that $Z$ is replaced by a multi-center warp-factor $Z_{1}$, is
that there is a source term for $G_{6}$ since we have introduced
D2 sources on this shell.

We can split both $H_{3}$ and $G_{6}$ into two pieces respectively:
\begin{eqnarray}
G_{6} &=&G_{6}^{(I)}+G_{6}^{(II)}, \\
H_{3} &=&H_{3}^{(I)}+g_{s}\ast _{9}G_{6}^{(II)},
\end{eqnarray}%
where $H_{3}^{(I)},G_{6}^{(I)}$ still satisfy the whole four equations (\ref%
{h3close})-(\ref{harmonic6}) with $Z$ replaced by the multi-center
$Z_{1}$. In large $r$ region when expanded around $r_{0}$,
the leading terms of $H_{3}^{(I)},G_{6}^{(I)}$ will reduce to the $n=-7$ solution in section 2. $H_{3}^{(I)},G_{6}^{(I)}$ are the contribution to the fluxes as if there were no D2 source. 

The influence of D2 source on the fluxes
are mainly on $G_{6}^{(II)}$, whose equations are now:
\begin{equation}
dG_{6}^{(II)}=J_{7},\quad \quad \quad \quad d\ast _{9}G_{6}^{(II)}=0.
\label{source-eqn}
\end{equation}
The contribution of $G_{6}^{(II)}$ is dominant over $G_{6}^{(I)}$ very close
to the shell since it has the delta function as source. Since $\ast
_{9}G_{6}^{(II)}$ is closed, it can be written as
\begin{equation}
\ast _{9}G_{6}^{(II)}=(r_{1}^{-2}\partial _{1}hdr_{1}+r_{1}^{-2}\partial
_{2}hdr_{2})\wedge \omega _{2},
\end{equation}%
where $\omega _{2}=r_{1}^{2}\cdot \mathrm{dvol}({S}^{2})$, and $%
\mathrm{dvol}({S}^{2})$ denotes the volume-form of an ${S}^{2}$
with unit radius embedded in 123 subspace and centered at origin, and $%
\partial _{1}\equiv \frac{\partial }{\partial r_{1}},$ $\partial _{2}\equiv
\frac{\partial }{\partial r_{2}}$. $h$ is a function of $r_{1},r_{2}$. The
solution (see appendix F for more detail) can be expressed through the function defined as $Y=r_{1}^{-2}\partial _{1}h,$ and we have
\begin{equation}
Y=\frac{4\pi C\kappa ^{2}\tau _{D2}g_{s}^{-2}pr_{0}^{2}}{5r_{1}}\partial
_{r_{0}}\left( r_{0}^{-1}\left[ \frac{1}{[(r_{1}+r_{0})^{2}+r_{2}^{2}]^{5/2}}%
-\frac{1}{[(r_{1}-r_{0})^{2}+r_{2}^{2}]^{5/2}}\right] \right) ,
\label{Y-result}
\end{equation}%
where the notations of the coefficients in $Y$ are in appendix F. 

The leading terms of $G_{6}^{(II)}$ and $H_{3}^{(II)}=g_{s}\ast _{9}G_{6}^{(II)}$ expanded in terms of $r_{0}$, in the region where $r_{1}\gg
r_{0}$, is precisely the $n=-9$ solution in section 2, and it is of third order in $r_{0}$ and thereby of third order in $\mu $ in large $r$ region
(see appendix F). The contribution of $G_{6}^{(II)}$ is dominant over
$G_{6}^{(I)}$ very close to the shell of the brane since it has
the delta function as source, while in large $r$ the situation is
reversed and $G_{6}^{(I)}$ becomes dominant over $G_{6}^{(II)}$
instead. We see the consistency in the calculation of the fluxes in the presence of the sources since the solutions of
$H_{3}^{(I)},G_{6}^{(I)}$ and $H_{3}^{(II)},G_{6}^{(II)}$ in this
section are the full solutions which just reduce to the $n=-7$ and
$n=-9$ solutions in large $r$ region in section 2 respectively.

\bigskip

\section{Metric and dilaton in large $r$ region and near each shell}

In this section, we come to the discussion on the metric and
dilaton. The situation is similar to Polchinski-Strassler
\cite{ps}\ that in most regions away from the shells, the
D0-charge dominates, and in the regions very close to the shells,
the D2-charge dominates instead. This switch of the role of the
dominance is also reflected in the change of the dominance between $
H_{3}^{(I)}, G_{6}^{(I)}$ and $H_{3}^{(II)}, G_{6}^{(II)}$ as discussed in
last section. We will discuss the metric and dilaton in two limiting regions
in this section. One is in the large $r$ region and the other is very near each
shell.

The general brane configuration in our solutions are concentric branes with
various D0-charge and radii. In the large $r$ region, since the D0-charge dominates, the
metric, dilaton and $F_{2}$ are very close to the near horizon
geometry of multi-center D0 branes with warp-factor $Z_{1}$
corresponding to the distributions of these concentric shells of
D0 branes. For the general configuration of several concentric
shells of ${S}^{2}$-wrapped branes with the $i$th shell having $p_{i}$ coincident D2 branes each with $q_{i}$ units of D0-charge ($N=\underset{i}{\sum
}p_{i}q_{i}$ and the D0-charge $q_{i}$ are all large and
distribute uniformly on the spheres), the warp factor $Z_{1}$ in the solution
of the near horizon geometry of multi-center D0 branes, as the
superposition of harmonic functions, should be
\begin{equation}
Z_{1}=\underset{i}{\sum }\frac{R_{i}^{7}}{10r_{1}r_{0}^{(i)}}\left[ \frac{1}{%
[(r_{1}-r_{0}^{(i)})^{2}+r_{2}^{2}]^{5/2}}-\frac{1}{%
[(r_{1}+r_{0}^{(i)})^{2}+r_{2}^{2}]^{5/2}}\right] ,  \label{z1manyshell}
\end{equation}%
where $R_{i}^{7}=60\pi ^{3}g_{s}(p_{i}q_{i})\alpha ^{^{\prime }7/2}$ and $%
r_{0}^{(i)}\approx \frac{\pi }{3}\widetilde{\mu }\alpha ^{^{\prime }}q_{i}$.

Now we will look at the metric and dilaton near
each shell of branes, say the $i$th shell. The total D0-charge of this shell is $N_{i}=p_{i}q_{i}$ and radius of this shell is $r_{0}^{(i)}\approx \frac{\pi }{3}\widetilde{\mu }\alpha ^{^{\prime }}q_{i}$. Very close to each shell, the D0-charge no longer have dominant influence
over D2-charge since the metric parallel to the shell expand and D0-charge
are diluted. We can approximate
the metric and dilaton near the $S^2$, e.g. without loss of generality,
near the point $(x_{1},x_{2},x_{3})=(0,0,r_{0}^{(i)})$, by the solution of $%
p_{i}$ flat D2 branes with $B_{2}$ potential on its spatial world-volume
\cite{noncomu}, \cite{noncomutative} (in string frame):
\begin{eqnarray}
ds^{2} &=&\frac{\alpha ^{^{\prime }5/2}u_{i}^{5/2}}{6\pi
^{2}g_{s}a_{i}^{5/2}p_{i}}\left( -dt^{2}+\frac{1}{1+a_{i}^{5}u_{i}^{5}}(d%
\widetilde{x}_{i1}^{2}+d\widetilde{x}_{i2}^{2})\right) +\frac{6\pi
^{2}g_{s}a_{i}^{5/2}p_{i}}{\alpha ^{^{\prime }1/2}u_{i}^{5/2}}%
(du_{i}^{2}+u_{i}^{2}d\Omega _{6}^{2}),  \nonumber \\
e^{2\Phi } &=&\frac{(6\pi ^{2})^{3}g_{s}^{5}a_{i}^{15/2}p_{i}^{3}}{\alpha
^{^{\prime }15/2}u_{i}^{15/2}}\left( \frac{a_{i}^{5}u_{i}^{5}}{%
1+a_{i}^{5}u_{i}^{5}}\right) ,  \label{flatd2b2}
\end{eqnarray}%
where the label $i$ denotes the $i$th shell. $\widetilde{x}_{i1}$,$\widetilde{x}_{i2}$ parameterize the spatial part of the D2 branes, $u_{i}$ is the energy direction
away from the $i$th shell of branes in the transverse direction and $a_{i}$ is a constant that will be worked out in (\ref{relation-of-parameter}). 

Since there is $B_{2}$ field on the D2 branes in solution (\ref{flatd2b2}), the D2 branes can couple to $C_{1}$ and there is D0-charge on the D2 branes. Suppose we are looking at regions only near the $i$th shell but not the other shells at the same time. When away from this
shell of branes such that$\ a_{i}^{5}u_{i}^{5}\gg 1,$ i.e. $\frac{%
a_{i}^{5}u_{i}^{5}}{1+a_{i}^{5}u_{i}^{5}}\approx 1$, the above metric and
dilaton (\ref{flatd2b2}) match exactly with the near horizon geometry of multi-center D0 branes, near this shell of branes ($x_{l}$: $x_{1},x_{2},x_{3}$ and $x_{a}$: $x_{4},...,x_{9}$):
\begin{eqnarray}
ds^{2} &=&-\frac{\sqrt{10}r_{0}^{(i)}\rho _{i}^{5/2}}{R_{i}^{7/2}}dt^{2}+%
\frac{R_{i}^{7/2}}{\sqrt{10}r_{0}^{(i)}\rho _{i}^{5/2}}(d\overrightarrow{x}%
_{l}^{2}+d\overrightarrow{x}_{a}^{2}),  \notag \\
e^{2\Phi } &=&g_{s}^{2}Z_{1}^{3/2}=\frac{g_{s}^{2}R_{i}^{21/2}}{%
10^{3/2}\left( r_{0}^{(i)}\right) ^{3}\rho _{i}^{15/2}},  \label{multiD0near}
\end{eqnarray}%
where $r_{0}^{(i)}\approx \frac{\pi }{3}\widetilde{\mu }\alpha ^{^{\prime
}}q_{i}$, $R_{i}^{7}=60\pi ^{3}g_{s}(p_{i}q_{i})\alpha ^{^{\prime }7/2}$. We
used the multi-center warp-factor $Z_{1}$ in (\ref{z1manyshell})
approximated near the $i$th shell of the brane source, and $\rho _{i}$ is
the distance away from the $i$th shell:
\begin{eqnarray}
Z_{1} &\approx &\frac{R_{i}^{7}}{10\left( r_{0}^{(i)}\right) ^{2}\rho
_{i}^{5}},  \label{z1nearshell} \\
\rho _{i} &=&[(r_{1}-r_{0}^{(i)})^{2}+r_{2}^{2}]^{1/2}.
\end{eqnarray}%

We can define a cross-over distance $\rho _{c}^{(i)}$ as the
distance away from the brane where $a_{i}u_{i}=1$, which could
characterize the regions of influence of the D2-charge. In order
for the match to be valid, we need when away
from the shell the two approximations $\frac{a_{i}^{5}u_{i}^{5}}{%
1+a_{i}^{5}u_{i}^{5}}\approx 1$, (i.e. $\rho _{i}\gg \rho _{c}^{(i)}$) and $%
Z_{1}\approx \frac{R_{i}^{7}}{10\left( r_{0}^{(i)}\right) ^{2}\rho _{i}^{5}}$%
, (i.e. $\rho _{i}\ll $ $r_{0}^{(i)}$) are both valid. This requires the
match to happen in the region $\rho _{c}^{(i)}\ll \rho _{i}\ll r_{0}^{(i)}$,
so for our approximation to be valid we need the parameters satisfy $\rho
_{c}^{(i)}$ $\ll r_{0}^{(i)}$.

Under this approximation, the dilaton and metric match exactly and we
thereby found the relation between the parameters and variables by comparing
the two solutions (\ref{flatd2b2}), (\ref{multiD0near}):
\begin{eqnarray}
u_{i} &=&\frac{\rho _{i}}{\alpha ^{^{\prime }}},\quad \quad a_{i}=\left(
\frac{R_{i}^{7/2}}{\sqrt{10}r_{0}^{(i)}\cdot 6\pi ^{2}g_{s}p_{i}}\right)
^{2/5},\quad \quad \rho _{c}^{(i)}=\alpha ^{^{\prime }}\left( \frac{\sqrt{10}%
r_{0}^{(i)}\cdot 6\pi ^{2}g_{s}p_{i}}{R_{i}^{7/2}}\right) ^{2/5},  \notag \\
\frac{\widetilde{x}_{i1,i2}}{x_{1,2}} &=&\frac{R_{i}^{7}}{10\left(
r_{0}^{(i)}\right) ^{2}\alpha ^{^{\prime }5/2}\cdot 6\pi ^{2}g_{s}p_{i}}.
\label{relation-of-parameter}
\end{eqnarray}%

So the metric and dilaton near the $i$th shell can be written as
\begin{eqnarray}
ds^{2} &=&-\frac{\sqrt{10}r_{0}^{{(i)}}\rho _{i}^{5/2}}{%
R_{i}^{7/2}}dt^{2}+\frac{R_{i}^{7/2}}{\sqrt{10}r_{0}^{{(i)}%
}\rho _{i}^{5/2}}(dx_{3}^{2}+d\overrightarrow{x}_{a}^{2})+\frac{%
R_{i}^{7/2}\rho _{i}^{5/2}}{\sqrt{10}r_{0}^{{(i)}}\left( \rho
_{i}^{5}+\left( \rho _{c}^{{(i)}}\right) ^{5}\right) }%
(dx_{1}^{2}+dx_{2}^{2}),  \notag \\
e^{2\Phi } &=&\frac{g_{s}^{2}R_{i}^{21/2}}{10^{3/2}\left( r_{0}^{%
{(i)}}\right) ^{3}\rho _{i}^{5/2}\left( \rho _{i}^{5}+\left(
\rho _{c}^{{(i)}}\right) ^{5}\right) },  \label{nearshell2}
\end{eqnarray}%
and as discussed above it is valid when
\begin{equation}
\frac{r_{0}^{(i)}}{\rho _{c}^{(i)}}\sim \left( \frac{q_{i}^{5}}{g_{s}N_{i}}%
\right) ^{1/5}\gg 1,  \label{chargecondition2}
\end{equation}%
which is just $\frac{q_{i}^{5}}{g_{s}N_{i}}\gg 1$, where $q_{i}$ is the
D0-charge of each D2 brane on the $i$th shell and $N_{i}$ is the total
D0-charge of the $i$th shell. This is the same scaling bound condition as (%
\ref{chargecondition}) in section 3.

For general brane configurations of several shells, the metric and
dilaton in complete regions are very difficult to solve
explicitly, but it's clear that in large $r$ region they approach the near horizon geometry of multi-center D0 branes with
warp-factor $Z_{1}$ in (\ref{z1manyshell}), and
near each shell they are approximated as the solutions in (\ref{nearshell2}%
). In special cases when there is only one single shell of branes, the metric and dilaton may be expressed approximately valid in all
regions. Suppose there are $p$ D2 branes each with D0-charge $q$ on this single
shell. $pq=N$ is the total D0 charge and the radius is $r_{0}\approx \frac{
\pi }{3}\widetilde{\mu }\alpha ^{^{\prime }}q$. The solution near the brane
can be obtained from (\ref{nearshell2}) by identifying $p_{i}=p,q_{i}=q$.
And then we can generalize the solution near the shell to all regions by
replacing the warp-factor $Z_{1}\approx \frac{R^{7}}{10r_{0}^{2}\rho ^{5}}
$, approximated near the shell as in (\ref{z1nearshell}), with the warp-factor $
Z_{1}$ in all regions as in (\ref{z1oneshell}). The validity of this approximation is again from (\ref{chargecondition2}), i.e. $\frac{q^{5}}{g_{s}N}\gg 1$. 

\bigskip

\section{Related issues and generalizations to other plane-wave M(atrix)
theories}

So far we have constructed the supergravity solutions dual to the
1/2 BPS concentric fuzzy sphere vacua of the BMN matrix model using the
method of the Polchinski-Strassler solution, which is the string dual of the
$\mathcal{N}=1^{\ast }$ theory. In this last section, we discuss some
related issues or remaining issues to our discussions above. 

Each 1/2 BPS vacuum of BMN matrix model can be represented by a Young
tableau \cite{transverse5}, \cite{protect} and it can be interpreted as
either concentric D2 branes or concentric NS5 branes, in different regimes
of parameters \cite{BMN perturb}, \cite{transverse5}. The configurations in terms of concentric shells of D2 branes
have their validities as dual descriptions when the effective 't Hooft
coupling in the matrix perturbation theory of the BMN matrix model when expanding around each fuzzy sphere is small\footnote{
The effective 't Hooft coupling is $p_{i}\left( \frac{1}{\mu p_{+}^{(i)}}%
\right) ^{3}$ \cite{BMN perturb}, where $p_{+}^{(i)}=\frac{q_{i}}{R}$ is the
light-cone momentum of the M2 brane on the $i$th shell and $p_{i}$ is the
number of coincident M2 branes on the $i$th shell.} \cite{BMN perturb}, \cite%
{transverse5}. When the effective 't Hooft couplings are small, the
interactions are smaller than the harmonic oscillator energies expanded
around these fuzzy spheres, and the BMN matrix model can be studied
perturbatively around these fuzzy spheres \cite{BMN perturb}. The concentric
D2 brane configurations are therefore good descriptions of the BMN vacua
when these parameters are small. For fixed partition of $N$, we can always
tune $\mu $ and $R$ to satisfy these conditions. For fixed $\mu $ and $R,$
it is relatively safer to expand around a fuzzy sphere when the numbers of
the coincident spheres are smaller and the matrix size of the sphere is
larger \cite{BMN perturb}.

We have analyzed the descriptions of the vacua in terms of concentric D2
branes in the regime of weak effective 't Hooft couplings and we haven't
studied the situation of the concentric NS5 branes in detail, which are
expected to be valid in the regime of strong effective 't Hooft couplings.
The NS5 branes polarized from D0 branes should also have equilibrium radii
with these additional fluxes turned on since it can couple to the dual of
the NS-NS 3-form flux and the 3-form potential via world-volume 3-form flux.

In fact, there could be smooth solutions that are dual to these vacua. The smooth solutions and the solutions with brane configurations studied in this paper may be related by geometric transitions (e.g. \cite{vafa}), where branes and fluxes get replaced with each other.

There are less supersymmetric vacua and time-dependent vacua e.g. \cite{BMN
perturb}, \cite{plefkaspectrum}, \cite{othervacua}, \cite{bak033}, \cite%
{shin}, \cite{instanton} that we have not discussed. For example,
it would be interesting to understand such as the 1/4 BPS rotating
elliptic fuzzy spheres described by \cite{bak033}, and the 1/4 BPS rotating fuzzy spheres by \cite{jeremy1}. There are also instanton solutions \cite%
{instanton} which are similar to the domain wall solutions in the $\mathcal{N%
}=1^{\ast }$ theory \cite{nahm}. It would be good to understand the dual
descriptions of the vacua in the model that is less supersymmetric and/or
non-static.

The Polchinski-Strassler type solution has been widely generalized and
applied to many other situations in terms of other branes (e.g. \cite%
{benaetal}, \cite{otherbrane}, \cite{dieletric}, \cite{inflation}). One can
conjecture that the Polchinski-Strassler type solution is universal for any
Dp branes \cite{puzzle}, which can be polarized to Dp+2 or NS5 branes in the
presence of the additional R-R and NS-NS fluxes on the background of the
near horizon geometry of $N$ Dp branes and the resulting solutions are dual
to the mass-deformed world-volume field theory of $N$ Dp branes. In each
such solution, we have a pair of R-R and NS-NS fluxes and this is mainly
because we have two channels of polarizations. The R-R flux is more
responsible for polarization to Dp+2 branes, while the NS-NS flux is more
responsible for polarization to NS5 branes.

The construction of dual supergravity descriptions to the BMN matrix model
may be generalized to those of the M(atrix) theories on less supersymmetric plane-wave backgrounds \cite{general-ppwave}, \cite{jeremy2}. For a general plane-wave matrix theory
with lagrangian, e.g. \cite{general-ppwave}, \cite{superp}:
\begin{equation*}
L=\frac{1}{2}\mathrm{Tr}\{\underset{i}{\sum }(D_{0}X^{i})^{2}+\underset{i,j}{%
\sum }\frac{1}{2}[X^{i},X^{j}]^{2}+i\psi ^{T}D_{0}\psi -\psi ^{T}\gamma
_{i}[X^{i},\psi ]
\end{equation*}%
\begin{equation}
-\underset{i}{\sum }\mu _{i}^{2}\left( X^{i}\right) ^{2}+\frac{2}{3}i%
\widetilde{T}_{ijk}X^{i}X^{j}X^{k}-\frac{1}{4}i\psi ^{T}\widetilde{\not{T}%
}\psi \},
\end{equation}%
where $\widetilde{\not{T}}{=}\frac{1}{3!}\widetilde{T}_{ijk}\gamma
^{ijk}$, $\underset{i}{\sum }\mu _{i}^{2}=\underset{i,j,k}{\sum }\frac{1}{12}%
(\widetilde{T}_{ijk})^{2}$, there might exist supergravity duals
which are similar to the case of the BMN matrix model. In the
large $r$ region, the dual supergravity solutions can also be considered
as perturbations around the near horizon geometry of $N$ D0 branes by
the additional fluxes $H_{3}$ and $G_{6}$, which should also
satisfy equations (\ref{h3close})-(\ref{harmonic6}), and the form
$Z^{-1}(H_{3}-g_{s}\ast _{9}G_{6})$, where
$Z=\frac{R^{7}}{r^{7}}$, should thereby also be annihilated by
both $d$ and $d\ast _{9}$ in the transverse 9-d. When $r$ goes to
infinity it should approach a constant form, so the additional
fluxes should satisfy the relation in large $r$ region as:
\begin{equation}
Z^{-1}(H_{3}-g_{s}\ast _{9}G_{6})\propto \widetilde{T}_{3},
\end{equation}%
where
$\widetilde{T}_{3}=\frac{1}{3!}\widetilde{T}_{ijk}dx^{i}\wedge
dx^{j}\wedge dx^{k}$, and $\widetilde{T}_{ijk}$ are the
coefficients of the Myers term in the corresponding M(atrix) theory
on the general plane-wave background. Thereby one can conjecture
that the perturbation by these mass terms and Myers terms are dual
to turning on these additional fluxes $H_{3}$ and $G_{6}$ causing
D0 branes to polarize into some non-spherical branes. The brane
configuration would be more difficult to describe than the case of
the BMN matrix model. For example, in the M(atrix) theory on the background of
T-dual of the IIB pp-wave lifted to 11d, the BPS vacua correspond to M2 branes polarized into M5 brane, where the M2 branes distributed on a fuzzy ellipsoid \cite{jeremy2}. So generally, the brane in the small $r$ region would take
the shape that corresponds to the classical static solution in the
corresponding M(atrix) theory and it should also equivalently be the
shape of a probe brane in the presence of external fluxes $H_{3}$
and $G_{6}.$ 

\section*{Acknowledgments}

I am very indebted to my advisor Juan Maldacena for sharing ideas
and for many discussions and encouragement. I am also very
grateful to Oleg Lunin and David Shih for commenting on the
manuscript and giving suggestions. The work was supported in part
by Princeton University Assistantship in Instruction.

\bigskip

\appendix

\section{IIA equations for fluxes}

The IIA equations of motion in this appendix are in Einstein frame and in
the convention of \cite{iia}. The bosonic equations of motion are:
\begin{eqnarray}
d\ast d(\Phi ) &=&-\frac{1}{2}g_{s}e^{-\Phi }H_{3}\wedge \ast H_{3}+\frac{3}{%
4}g_{s}^{1/2}e^{3\Phi /2}F_{2}\wedge \ast F_{2}+\frac{1}{4}%
g_{s}^{3/2}e^{\Phi /2}\widetilde{F}_{4}\wedge \ast \widetilde{F}_{4},
\label{dilatoneqn} \\
d(e^{3\Phi /2}\ast F_{2}) &=&g_{s}e^{\Phi /2}H_{3}\wedge \ast \widetilde{F}%
_{4},  \label{f2eqn} \\
d(e^{\Phi /2}\ast \widetilde{F}_{4}) &=&-g_{s}^{1/2}F_{4}\wedge H_{3},
\label{f4eqn} \\
\frac{1}{2}g_{s}F_{4}\wedge F_{4} &=&d(e^{-\Phi }\ast
H_{3}+g_{s}^{1/2}e^{\Phi /2}C_{1}\wedge \ast \widetilde{F}_{4}),
\label{h3eqn}
\end{eqnarray}%
where $F_{2}=dC_{1},F_{4}=dC_{3},H_{3}=dB_{2},\widetilde{F}%
_{4}=F_{4}-C_{1}\wedge H_{3},$ and the Bianchi identities are: $%
dF_{2}=0,dF_{4}=0,dH_{3}=0.$ The relation between the Einstein frame metric
and string frame metric is $\left( G_{\mu \nu }\right) _{\mathrm{Einstein}%
}=g_{s}^{1/2}e^{-\Phi /2}\left( G_{\mu \nu }\right) _{\mathrm{string}}$.

The unperturbed background is the near horizon geometry of $N$ D0 branes. In
Einstein frame it is
\begin{eqnarray}
ds^{2} &=&-Z^{-7/8}dt^{2}+Z^{1/8}d\overrightarrow{x}_{i}^{2},\quad \quad
i=1,...,9,  \notag \\
e^{\Phi } &=&g_{s}Z^{3/4},\quad \quad C_{1}=g_{s}^{-1}(Z^{-1}-1)dt,\quad
\quad Z=\frac{R^{7}}{r^{7}}.
\end{eqnarray}

The perturbed fluxes $H_{3}$ and $\widetilde{F}_{4}$ are small fluctuations
in large $r$ region. The perturbed terms in the right sides of equation (\ref%
{dilatoneqn}), (\ref{f2eqn}), if non-zero, are at least of second order
fluctuations. Therefore if neglecting the terms of second or higher orders
in the fluctuations, the equations for the dialton and $F_{2}$ still have
the form:
\begin{equation}
d\ast d(\Phi )=\frac{3}{4}g_{s}^{1/2}e^{3\Phi /2}F_{2}\wedge \ast
F_{2},\quad \quad \quad d(e^{3\Phi /2}\ast F_{2})=0,\quad \quad \quad
dF_{2}=0.
\end{equation}%
And therefore the other fluxes $H_{3}$ ,$\widetilde{F}_{4}$ obey:
\begin{eqnarray}
\quad \quad \quad d(Z^{3/8}\ast \widetilde{F}_{4}) &=&0,\quad \quad \quad
dH_{3}=0.  \label{closeeqn2} \\
dF_{4} &=&0,\quad \quad \quad d(g_{s}^{-1}Z^{-3/4}\ast
H_{3}+g_{s}Z^{3/8}C_{1}\wedge \ast \widetilde{F}_{4})=0.
\label{harmoniceqn2}
\end{eqnarray}

From (\ref{closeeqn2}) we can define $G_{6}=Z^{3/8}\ast \widetilde{F}_{4},$
so that $dG_{6}=0.$ We can write (\ref{harmoniceqn2}) in terms of the 9-d
Hodge dual $\ast _{9}$ in the transverse 9-d with respect to the flat
metric: $\ast H_{3}=dx^{0}\wedge (-Z^{-1/4}\ast _{9}H_{3})$, $\widetilde{%
F_{4}}=dx^{0}\wedge (-Z^{-1}\ast _{9}G_{6})$, then from (\ref{harmoniceqn2})
we precisely arrive at equation (\ref{harmonic3}), (\ref{harmonic6}) in
section 2.

\section{Linearized solutions of fluxes}

In this appendix we discuss linearized solutions of additional fluxes in
terms of tensor harmonics. Without loss of generality, we can define a $%
T_{3} $ analogous to Polchinski-Strassler \cite{ps} as follows:
\begin{eqnarray}
T_{3} &=&dr_{1}\wedge \omega _{2}=dx^{1}\wedge dx^{2}\wedge dx^{3}, \\
S_{2} &=&r_{1}\omega _{2}=\frac{1}{2!}\varepsilon _{ijk}x^{i}\wedge
dx^{j}\wedge dx^{k},\quad \quad dS_{2}=3T_{3}, \\
V_{3} &=&d\ln r\wedge S_{2}=\frac{r_{1}^{2}}{r^{2}}dr_{1}\wedge \omega _{2}+%
\frac{r_{1}r_{2}}{r^{2}}dr_{2}\wedge \omega _{2} \\
&=&\frac{r_{1}^{2}}{r^{2}}dx^{1}\wedge dx^{2}\wedge dx^{3}+\frac{1}{2!}\frac{%
x^{a}x^{i}}{r^{2}}\varepsilon _{ijk}dx^{a}\wedge dx^{j}\wedge dx^{k}, \\
\quad \quad (i,j,k &=&1,2,3,a=4,...,9)  \notag
\end{eqnarray}%
where the definition of $\omega _{2}$, $\omega _{5}$ are in section 4. Then
\begin{eqnarray}
\quad \quad \ast _{9}T_{3} &=&dr_{2}\wedge \omega _{5}, \\
\ast _{9}V_{3} &=&\frac{r_{1}^{2}}{r^{2}}dr_{2}\wedge \omega _{5}-\frac{%
r_{1}r_{2}}{r^{2}}dr_{1}\wedge \omega _{5}.
\end{eqnarray}

If we search for the solutions of the form in large $r$ region according to
the isometry $SO(3)\times SO(6)$:
\begin{equation}
\quad H_{3}=r^{m}(\alpha T_{3}+\beta V_{3}),\quad \quad \quad \quad
G_{6}=g_{s}^{-1}r^{n}(\gamma \ast _{9}T_{3}+\delta \ast _{9}V_{3}),
\end{equation}%
where $m,n,\alpha ,\beta ,\gamma ,\delta $ are constants. The $g_{s}^{-1}$
factor in the ansatz for $G_{6}$ is introduced for convenience in this
appendix. The four equations (\ref{h3close}), (\ref{g6close}), (\ref%
{harmonic3}), (\ref{harmonic6}) give the constraints:
\begin{equation}
dH_{3}=0,\quad \Rightarrow \quad \beta =\frac{m}{3}\alpha .\quad
\end{equation}%
\begin{equation}
dG_{6}=0,\quad \Rightarrow \quad \delta =-\frac{n}{n+6}\gamma .
\end{equation}%
\begin{equation}
d[Z^{-1}(H_{3}-g_{s}\ast _{9}G_{6})]=0,\quad \Rightarrow \quad 7\alpha r^{m}-%
\frac{n^{2}+16n+42}{n+6}\gamma r^{n}=0.
\end{equation}%
\begin{equation}
d[Z^{-1}(\ast _{9}H_{3}-g_{s}G_{6})]=0,\quad \Rightarrow \quad \frac{%
m^{2}+16m+21}{3}\alpha r^{m}-\frac{42}{n+6}\gamma r^{n}=0.
\end{equation}

Since in the perturbation, $\alpha ,\gamma $ cannot be both zero, we need $%
m,n$ to be equal, and then we get
\begin{equation}
n(n+16)(n+7)(n+9)=0.
\end{equation}

This leads to 4 solutions that are two pairs of non-normalizable and
normalizable solutions in (\ref{n=-7}), (\ref{n=-9}), (\ref{n=0}), (\ref%
{n=-12}) in section 2.

\section{Killing spinors}

The IIA susy transformation rules we used are in string frame (e.g. \cite%
{susyrule}):
\begin{eqnarray}
\delta \psi _{m} &=&[(\partial _{m}+\frac{1}{4}\omega _{mab}\Gamma ^{ab}+%
\frac{1}{8}e^{\Phi }\not{F}_{2}\Gamma _{m}\Gamma ^{\underline{11}})+(%
\frac{1}{8}H_{mab}\Gamma ^{ab}\Gamma ^{\underline{11}}+\frac{1}{8}e^{\Phi }%
\widetilde{\not{F}}_{4}\Gamma _{m})]\epsilon , \\
\delta \lambda &=&[(\frac{1}{2}\Gamma ^{m}\partial _{m}\Phi +\frac{3}{8}%
e^{\Phi }\not{F}_{2}\Gamma ^{\underline{11}})+(\frac{1}{4}\not{H}%
_{3}\Gamma ^{\underline{11}}+\frac{1}{8}e^{\Phi }\widetilde{\not{F}}%
_{4})]\epsilon ,
\end{eqnarray}%
where letters with a slash denote the contractions of forms with gamma matrices: $\not{F}_{2}=\frac{1}{2!}F_{ab}\Gamma ^{ab},{\not{H}}_{3}=\frac{1}{3!}%
H_{abc}\Gamma ^{abc},\widetilde{\not{F}}_{4}=\frac{1}{4!}\widetilde{F}%
_{abcd}\Gamma ^{abcd}$ and similarly for other forms.

The unperturbed Killing spinors $\epsilon ^{(0)}$\ in the absence of $%
H_{3},G_{6}$\ satisfy:
\begin{eqnarray}
\lbrack \frac{1}{2}\Gamma ^{m}\partial _{m}\Phi +\frac{3}{8}e^{\Phi }\not{%
F}_{2}\Gamma ^{\underline{11}}]\epsilon ^{(0)} &=&\frac{3}{8}Z^{-5/4}\Gamma
^{\underline{i}}\partial _{i}Z(1+\Gamma ^{\underline{0}}\Gamma ^{\underline{%
11}})\epsilon ^{(0)}=0, \\
\lbrack \partial _{0}+\frac{1}{4}\omega _{0ab}\Gamma ^{ab}+\frac{1}{8}%
e^{\Phi }\not{F}_{2}\Gamma _{0}\Gamma ^{\underline{11}}]\epsilon ^{(0)}
&=&[\partial _{0}+\frac{1}{8}Z^{-3/2}\Gamma ^{\underline{0}}\Gamma ^{%
\underline{i}}\partial _{i}Z(1+\Gamma ^{\underline{0}}\Gamma ^{\underline{11}%
})]\epsilon ^{(0)}=0, \\
\lbrack \partial _{i}+\frac{1}{4}\omega _{iab}\Gamma ^{ab}+\frac{1}{8}%
e^{\Phi }\not{F}_{2}\Gamma _{i}\Gamma ^{\underline{11}}]\epsilon ^{(0)}
&=&[(\partial _{i}+\frac{1}{8}Z^{-1}\partial _{i}Z)-\frac{1}{8}%
Z^{-1}\partial _{i}Z(1+\Gamma ^{\underline{0}}\Gamma ^{\underline{11}})
\notag \\
&&+\frac{1}{8}Z^{-1}\Gamma ^{\underline{i}}\Gamma ^{\underline{j}}\partial
_{j}Z(1+\Gamma ^{\underline{0}}\Gamma ^{\underline{11}})]\epsilon ^{(0)}=0.\
(j\neq i),\quad \quad
\end{eqnarray}%
where the indices $i,j$ denotes 1,...,9 and the indices of gamma matrices
with a bar below are the gamma matrices in tangent space. So we have the
already familiar result (e.g. \cite{g=1}, \cite{d0multipole}):
\begin{equation}
\epsilon ^{(0)}=Z^{-1/8}\eta ,  \label{ep0}
\end{equation}%
where $\eta $\ is a constant spinor satisfying $(1+\Gamma ^{\underline{0}%
}\Gamma ^{\underline{11}})\eta =0.$\ The Killing spinors in the presence of
small fluctuations of $H_{3},G_{6}$\ can be written as$\ $%
\begin{equation}
\epsilon =\epsilon ^{(0)}+\epsilon ^{(1)},
\end{equation}%
where $\epsilon ^{(1)}$\ is the perturbation around $\epsilon ^{(0)}$. When
we turn on the $H_{3},G_{6}$\ that are of first order in $\mu $, $\epsilon
^{(1)}$\ is of order $\mu$. The variations of gravitino and dilatino of
the first order in\ $\mu $\ give:
\begin{eqnarray}
\frac{3}{8}Z^{-5/4}\Gamma ^{\underline{i}}\partial _{i}Z(1+\Gamma ^{%
\underline{0}}\Gamma ^{\underline{11}})\epsilon ^{(1)} &=&-[\frac{1}{4}%
\not{H}_{3}\Gamma ^{\underline{11}}+\frac{1}{8}e^{\Phi }\widetilde{%
\not{F}}_{4}]\epsilon ^{(0)},  \label{dilatino2} \\
\lbrack \partial _{0}+\frac{1}{8}Z^{-3/2}\Gamma ^{\underline{0}}\Gamma ^{%
\underline{i}}\partial _{i}Z(1+\Gamma ^{\underline{0}}\Gamma ^{\underline{11}%
})]\epsilon ^{(1)} &=&-[\frac{1}{8}H_{0ab}\Gamma ^{ab}\Gamma ^{\underline{11}%
}+\frac{1}{8}e^{\Phi }\widetilde{\not{F}}_{4}\Gamma _{0}]\epsilon
^{(0)},\   \label{gravitinotime2}
\end{eqnarray}%
\begin{eqnarray}
&&[(\partial _{i}+\frac{1}{8}Z^{-1}\partial _{i}Z)-\frac{1}{8}Z^{-1}\partial
_{i}Z(1+\Gamma ^{\underline{0}}\Gamma ^{\underline{11}})+\frac{1}{8}%
Z^{-1}\Gamma ^{\underline{i}}\Gamma ^{\underline{j}}\partial _{j}Z(1+\Gamma
^{\underline{0}}\Gamma ^{\underline{11}})]\epsilon ^{(1)}  \notag \\
&=&-[\frac{1}{8}H_{iab}\Gamma ^{ab}\Gamma ^{\underline{11}}+\frac{1}{8}%
e^{\Phi }\widetilde{\not{F}}_{4}\Gamma _{i}]\epsilon ^{(0)}.\ (j\neq i.)
\label{gravitinospace2}
\end{eqnarray}%
Now we can first consider the dilatino variation (\ref{dilatino2}) involving
$\epsilon ^{(1)}$. Since $\epsilon ^{(0)}$\ is not time-dependent while $%
\epsilon ^{(1)}$\ is time-dependent, the time-dependent part of $\epsilon
^{(1)}$\ should be annihilated by $(1+\Gamma ^{\underline{0}}\Gamma ^{%
\underline{11}})$. So we can decompose $\epsilon ^{(1)}$\ into two parts: $\
\epsilon ^{(1)}=$\ $\epsilon _{1}^{(1)}+$\ $\epsilon _{2}^{(1)}$, where $%
\epsilon _{1}^{(1)}$\ is time-dependent and $\epsilon _{2}^{(1)}$\ is not
time-dependent.

We can then split dilatino equation (\ref{dilatino2}) into two equations:%
\begin{eqnarray}
(1+\Gamma ^{\underline{0}}\Gamma ^{\underline{11}})\epsilon _{1}^{(1)} &=&0,
\label{ep1} \\
\Gamma ^{\underline{i}}\partial _{i}Z(1+\Gamma ^{\underline{0}}\Gamma ^{%
\underline{11}})\epsilon _{2}^{(1)} &=&-\frac{8}{3}Z^{5/4}[\frac{1}{4}%
\not{H}_{3}\Gamma ^{\underline{11}}+\frac{1}{8}e^{\Phi }\widetilde{%
\not{F}}_{4}]\epsilon ^{(0)}.  \label{ep2}
\end{eqnarray}

Substituting (\ref{ep1}), (\ref{ep2}) into the time-component of the
gravitino variation, equation (\ref{gravitinotime2}), and the spatial
components of gravitino variation, equation (\ref{gravitinospace2}), we
have:
\begin{equation}
\partial _{0}\epsilon _{1}^{(1)}=\frac{1}{12}Z^{-1/4}\Gamma ^{\underline{0}}[%
\not{H}_{3}\Gamma ^{\underline{11}}-e^{\Phi }\widetilde{\not{F}}%
_{4}]\epsilon ^{(0)},\quad \quad \quad (\partial _{i}+\frac{1}{8}%
Z^{-1}\partial _{i}Z)\epsilon _{1}^{(1)}=0.  \label{ep1dt}
\end{equation}

Since the right side of the first equation in (\ref{ep1dt}) is
time-independent, we solve that $\epsilon _{1}^{(1)}$\ is linear in time:
\begin{equation}
\epsilon _{1}^{(1)}=\frac{1}{12}\cdot \frac{1}{3!}Z^{-1}[H_{ijk}-g_{s}(\ast
_{9}G_{6})_{ijk}]\Gamma ^{\underline{i}\underline{j}\underline{k}}\epsilon
^{(0)}t,
\end{equation}%
and $\epsilon _{1}^{(1)}$ has the same helicity to $\epsilon ^{(0)}$. The
spatial-independence of $Z^{-1/8}\epsilon _{1}^{(1)}$ from the second
equation in (\ref{ep1dt}) and (\ref{ep0}) imply:
\begin{equation}
Z^{-1}[H_{ijk}-g_{s}(\ast _{9}G_{6})_{ijk}]=\mathrm{const.}  \label{constant}
\end{equation}%
The discussion so far doesn't require $SO(3)\times SO(6)$\ symmetry but only
that the fluxes $H_{3}$ and $G_{6}$ be small. There are stronger constraints
than merely that $Z^{-1}[H_{3}-g_{s}\ast _{9}G_{6}]$\ would be a constant
from the spatial part of the gravitino variation involving $\epsilon
_{2}^{(1)}$ from (\ref{gravitinospace2}):
\begin{eqnarray}
&&[(\partial _{i}+\frac{1}{8}Z^{-1}\partial _{i}Z)-\frac{1}{8}Z^{-1}\partial
_{i}Z(1+\Gamma ^{\underline{0}}\Gamma ^{\underline{11}})+\frac{1}{8}%
Z^{-1}\Gamma ^{\underline{i}}\Gamma ^{\underline{j}}\partial _{j}Z(1+\Gamma
^{\underline{0}}\Gamma ^{\underline{11}})]\epsilon _{2}^{(1)}  \notag \\
&=&[-\frac{1}{8}H_{iab}\Gamma ^{ab}+\frac{1}{8}Z^{1/4}\not{G}_{3}\Gamma ^{%
\underline{i}}]\Gamma ^{\underline{0}}\epsilon ^{(0)},  \label{ep2ds}
\end{eqnarray}%
where we define $G_{3}=g_{s}\ast _{9}G_{6}$. If contracting both sides of
equation (\ref{ep2ds}) with $\Gamma ^{\underline{i}},$\ and input equation (%
\ref{ep2}), and then\ acting on both sides the projection $(1-\Gamma ^{%
\underline{0}}\Gamma ^{\underline{11}}),$\ the right side becomes zero and
we have
\begin{equation}
(1-\Gamma ^{\underline{0}}\Gamma ^{\underline{11}})\epsilon _{2}^{(1)}=0.
\label{eps2projection}
\end{equation}%
So $\epsilon _{2}^{(1)}$ has the opposite chirality with respect to $%
\epsilon ^{(0)}$ and $\epsilon _{1}^{(1)}.$

From now on, we will use the indices $l,m,n$\ to denote 1,2,3,4,...,9, the
indices $i,j,k$\ to denote 1,2,3, and the\ indices $a,b,c$\ to denote
4,...,9, for convenience. Applying the projection condition (\ref%
{eps2projection}) on $\epsilon _{2}^{(1)}$ and then equation (\ref{ep2})
becomes:
\begin{equation}
\Gamma ^{\underline{l}}\partial _{l}Z\epsilon _{2}^{(1)}=-\frac{4}{3}Z^{5/4}%
\left[ \frac{1}{4}\not{H}_{3}+\frac{1}{8}\not{G}_{3}\right] \Gamma ^{%
\underline{0}}\epsilon ^{(0)}.\   \label{ep2contract}
\end{equation}

Substituting (\ref{ep2contract}) into (\ref{ep2ds}) and using the projection
condition (\ref{eps2projection}), we have 9 individual spatial equations:
\begin{equation}
\partial _{l}[Z^{-3/8}\epsilon _{2}^{(1)}]=Z^{-1/4}\left[ \frac{1}{12}\Gamma
^{\underline{l}}\not{H}_{3}-\frac{1}{8}H_{\underline{l}\underline{m}%
\underline{n}}\Gamma ^{\underline{m}\underline{n}}+\frac{1}{24}\Gamma ^{%
\underline{l}}\not{G}_{3}+\frac{1}{8}\not{G}_{3}\Gamma ^{\underline{l}}%
\right] \Gamma ^{\underline{0}}\eta .  \label{ep2ds2}
\end{equation}

Now if the we look at the fluxes in the ansatz (\ref{flux-ansat-h3}), (\ref%
{flux-ansat-g6})\ and combine the first supersymmetry condition from (\ref%
{constant}), we have
\begin{equation}
H_{3}=R^{7}r^{-7}(-\alpha T_{3}-\beta V_{3}),\quad \quad \quad
G_{3}=R^{7}r^{-7}(\gamma T_{3}-\beta V_{3}),
\end{equation}%
where $\alpha ,\beta ,\gamma $ are constants.

By dimensional analysis from equation (\ref{ep2contract}), (\ref{ep2ds2}),
one finds that\ $\epsilon _{2}^{(1)}$ after extracted out the$\ Z^{3/8}$
factor should be linear in $x^{l}$ so we can try the ansatz:
\begin{equation}
\epsilon _{2}^{(1)}=(\mu _{1}\Gamma ^{\underline{i}}x^{i}+\mu _{4}\Gamma ^{%
\underline{a}}x^{a})Z^{3/8}\Gamma ^{\underline{1}\underline{2}\underline{3}%
}\Gamma ^{\underline{0}}\eta ,
\end{equation}%
where $\mu _{1},$ $\mu _{4}$ are constants. Then comparing the left and
right side of equation (\ref{ep2contract}), we get the relation
\begin{equation}
\mu _{1}=\frac{-2\alpha +\gamma }{42}-\frac{\beta }{14},\quad \quad \quad
\mu _{4}=\frac{-2\alpha +\gamma }{42}.  \label{parameter1}
\end{equation}%
Comparing the left and right side of equation (\ref{ep2ds2}), we get another
relation
\begin{equation}
\mu _{1}=\frac{\alpha +\gamma }{6},\quad \quad \quad \mu _{4}=-\frac{\alpha
+\gamma }{12}.  \label{parameter2}
\end{equation}%
This shows $\mu _{1}:$\ $\mu _{4}=2:-1,$ and combine with (\ref{parameter1})
and (\ref{parameter2}) we have $\alpha :\beta :\gamma =3:-7:-1$, which is
just our $n=-7$ solution (\ref{n=-7flux}) in section 2.

\section{Approximation of D2 potential}

In this appendix, we write some details in the approximation of D2 potential in section 3. The DBI and WZ
action of the a D2 brane with $q$ units of D0 charge in the string frame is
in (\ref{d2action})
\begin{equation}
S_{D2}=-\tau _{D2}\int d^{3}\sigma e^{-\Phi }\sqrt{-\det (G_{\alpha \beta
}+2\pi \alpha ^{^{\prime }}\mathcal{F}_{\alpha \beta })}+\tau _{D2}\int
(C_{3}+2\pi \alpha ^{^{\prime }}\mathcal{F}_{2}\wedge C_{1}),
\end{equation}%
where $2\pi \alpha ^{^{\prime }}\mathcal{F}_{2}=2\pi \alpha ^{^{\prime
}}F_{2}-B_{2}$. We choose the gauge that the world-volume coordinates are
the same as the space-time ones, i.e. $t,\theta ,\varphi ,$ where the angles
parameterize the ${S}^{2}$. $G_{\shortparallel },G_{\perp }$ are the
pull-back metrics parallel to the time and the spherical directions
respectively, so we have:
\begin{equation}
\det G_{\shortparallel }=-Z^{-1/2},\quad \quad \quad \det G_{\perp
}=Zr_{1}^{4}\sin ^{2}\theta ,
\end{equation}%
where $r_{1}$ is the radius in 123 subspace.

The D0-charge of the D2 brane is $q$, so the world-volume 2-form fluxes $%
F_{2}$ is:
\begin{equation}
F_{2}=\frac{1}{2}q\sin \theta d\theta \wedge d\varphi ,\quad \quad \quad
\underset{S^{2}}{\int }F_{2}=2\pi q,
\end{equation}%
so we have $F_{\theta \varphi }=\frac{1}{2}q\sin \theta $, $\det F_{2}=\frac{%
1}{2}F_{\alpha \beta }F^{\alpha \beta }\det G_{\perp }=\frac{q^{2}}{4}\sin
^{2}\theta $, $F_{\alpha \beta }F^{\alpha \beta }=\frac{q^{2}}{2Zr_{1}^{4}}$.

Suppose the dominating terms in both DBI and WZ actions are from the
contribution of $F_{2}$, which requires the conditions (\ref{condition1})
then we can expand the square-root in the DBI action as
\begin{equation}
\sqrt{-\det (G_{\alpha \beta }+2\pi \alpha ^{^{\prime }}\mathcal{F}_{\alpha
\beta })}\approx \sqrt{-\det G_{\shortparallel }}\left( 2\pi \alpha
^{^{\prime }}\sqrt{\det F_{2}}+\frac{\det G_{\perp }}{4\pi \alpha ^{^{\prime
}}\sqrt{\det F_{2}}}\right) .
\end{equation}

The leading term in the DBI part is%
\begin{equation}
-\tau _{D2}g_{s}^{-1}\int d^{3}\sigma Z^{-3/4}\sqrt{-\det G_{\shortparallel }%
}\cdot 2\pi \alpha ^{^{\prime }}\sqrt{\det F_{2}}=-\int dtd\theta d\varphi
\left( 2\pi \alpha ^{^{\prime }}\tau _{D2}g_{s}^{-1}Z^{-1}\cdot \frac{1}{2}%
q\sin \theta \right) .
\end{equation}

The leading term in the WZ part is
\begin{equation}
\tau _{D2}\int 2\pi \alpha ^{^{\prime }}F_{2}\wedge C_{1}=\int dtd\theta
d\varphi \left( 2\pi \alpha ^{^{\prime }}\tau _{D2}g_{s}^{-1}Z^{-1}\cdot
\frac{1}{2}q\sin \theta \right) ,
\end{equation}%
where we choose the gauge choice $C_{1}=g_{s}^{-1}Z^{-1}dx^{0}$ in section
3. The two leading terms from the DBI part and WZ part precisely cancel.

The subleading terms of the DBI and WZ parts read respectively:
\begin{equation}
-\tau _{D2}g_{s}^{-1}\int d^{3}\sigma Z^{-3/4}\frac{\sqrt{-\det
G_{\shortparallel }}\det G_{\perp }}{4\pi \alpha ^{^{\prime }}\sqrt{\det
F_{2}}}=-\int dtd\theta d\varphi \left( \tau _{D2}g_{s}^{-1}\frac{%
r_{1}^{4}\sin \theta }{2\pi \alpha ^{^{\prime }}q}\right) =\int dt\left( -%
\frac{2\tau _{D2}r_{1}^{4}}{g_{s}\alpha ^{^{\prime }}q}\right) ,
\label{DBI subleading}
\end{equation}%
where the $Z$ factor cancel exactly in (\ref{DBI subleading}) and%
\begin{equation}
\tau _{D2}\int C_{3}=\int dtd\theta d\varphi \left( \frac{1}{3}\tau
_{D2}g_{s}^{-1}\widetilde{\mu }r_{1}^{3}\sin \theta \right) =\int dt\left(
\frac{4\pi \tau _{D2}\widetilde{\mu }r_{1}^{3}}{3g_{s}}\right) ,
\end{equation}%
where $C_{3}=\frac{1}{3}g_{s}^{-1}\widetilde{\mu }dx^{0}\wedge S_{2}$ since $%
F_{4}=\widetilde{F}_{4}+C_{1}\wedge H_{3}=-g_{s}^{-1}\widetilde{\mu }%
dx^{0}\wedge T_{3}$.

\section{Putting source}

The IIA bosonic\ action without external source is \cite{iia}
\begin{eqnarray}
S_{IIA} &=&\frac{1}{2\kappa ^{2}}[\int d^{10}x\sqrt{-G}R-\frac{1}{2}\int
(d\Phi \wedge \ast d\Phi +g_{s}e^{-\Phi }H_{3}\wedge \ast
H_{3}+g_{s}^{1/2}e^{3\Phi /2}F_{2}\wedge \ast F_{2}  \notag \\
&&+g_{s}^{3/2}e^{\Phi /2}\widetilde{F}_{4}\wedge \ast \widetilde{F}%
_{4}+g_{s}^{2}B_{2}\wedge F_{4}\wedge F_{4})].
\end{eqnarray}%
When we add D2 branes with D0-charge distributed in a single shell as in
section 4, we actually introduced the source terms in the action. In analogy
to electricity and magnetism, now in the action there should appear a source
term (some related discussion on adding source term is in \cite{addsource}):
\begin{equation}
S_{source}=\frac{1}{2\kappa ^{2}}\int g_{s}^{2}(C_{3}-B_{2}\wedge
C_{1})\wedge J_{7},
\end{equation}%
where $J_{7}=2\kappa ^{2}\tau _{D2}g_{s}^{-2}p\delta (r_{1}-r_{0})\delta
^{6}(r_{2})dr_{1}\wedge dr_{2}\wedge \omega _{5}$ and $\delta
^{6}(r_{2})=\delta (x_{4})\delta (x_{5})\delta (x_{6})\delta (x_{7})\delta
(x_{8})\delta (x_{9})$. The coefficient in $J_{7}$ are read off from
comparing the WZ action of the D2 branes with D0-charge involving the
couplings to $C_{3}$, $B_{2}$. So now the total action is
\begin{equation}
S=S_{IIA}+S_{source}.
\end{equation}

Since now $C_{3}$\ couples to $J_{7},$ and $B_{2}$ couples to $\frac{1}{2}%
F_{4}\wedge F_{4}+C_{1}\wedge J_{7}$, the equations for $d(e^{\Phi /2}\ast
\widetilde{F}_{4})$\ and for $d(e^{-\Phi }\ast H_{3})$\ when adding source
should be modified to
\begin{eqnarray}
d(e^{\Phi /2}\ast \widetilde{F}_{4}) &=&-g_{s}^{1/2}F_{4}\wedge
H_{3}+g_{s}^{1/2}J_{7}, \\
d(e^{-\Phi }\ast H_{3}+g_{s}^{1/2}e^{\Phi /2}C_{1}\wedge \ast \widetilde{F}%
_{4}) &=&\frac{1}{2}g_{s}F_{4}\wedge F_{4}+g_{s}C_{1}\wedge J_{7}.
\end{eqnarray}%
In our cases, since the forms $H_{3}$ and $F_{4}$ both have overall factors
of the volume-form of the ${S}^{2}$ in 123 subspace due to the
isometry $SO(3)\times SO(6)$, the terms $F_{4}\wedge F_{4}$ and $F_{4}\wedge
H_{3}$ are zero. The equations of motion for the fluxes on the background of
the near horizon geometry of a shell of multi-center D0 branes with D2
sources turned on are then modified to:
\begin{eqnarray}
dH_{3} &=&0, \\
dG_{6} &=&J_{7}, \\
d[Z_{1}^{-1}(H_{3}-g_{s}\ast _{9}G_{6})] &=&0, \\
dx^{0}\wedge d[-g_{s}^{-1}Z_{1}^{-1}\ast _{9}H_{3}+(Z_{1}^{-1}-1)G_{6}]
&=&g_{s}C_{1}\wedge J_{7}.  \label{harmonic6-modify2}
\end{eqnarray}

Since $J_{7}$\ is a delta function located at the ${S}^{2}$\ with
radius $r_{0}$, so for the right side of equation (\ref{harmonic6-modify2}),
we need to consider the $C_{1}$\ at $r_{1}=r_{0}$, $r_{2}=0.$\ The
multi-center warp-factor $Z_{1}$\ on the location of the delta function is
infinity,\ so $C_{1}=-g_{s}^{-1}dx^{0}$\ on the location of the delta
function, where we used the gauge choice $%
C_{1}=g_{s}^{-1}(Z_{1}^{-1}-1)dx^{0}$. Since $dG_{6}=J_{7}$, the last
equation (\ref{harmonic6-modify2}) becomes $d[Z_{1}^{-1}(\ast
_{9}H_{3}-g_{s}G_{6})]=0$.

\section{Solving the equation with source}

In this appendix we solve the equations in (\ref{source-eqn}) in section 4.
Since $\ast _{9}G_{6}^{(II)}$ is closed, it can be written as $\ast
_{9}G_{6}^{(II)}=(r_{1}^{-2}\partial _{1}hdr_{1}+r_{1}^{-2}\partial
_{2}hdr_{2})\wedge \omega _{2},$ so the equation $dG_{6}^{(II)}=J_{7}$ gives
\begin{equation}
\partial _{1}^{2}h-2r_{1}^{-1}\partial _{1}h+\partial
_{2}^{2}h+5r_{2}^{-1}\partial _{2}h=2\kappa ^{2}\tau
_{D2}g_{s}^{-2}pr_{0}^{2}\delta (r_{1}-r_{0})\delta ^{6}(r_{2}).
\label{pre-laplacian}
\end{equation}%
We can convert equation (\ref{pre-laplacian}) to a Laplacian equation with
source terms by making a derivative with respect to $r_{1}$ of both sides of
equation (\ref{pre-laplacian}) and define the function $Y=r_{1}^{-2}\partial
_{1}h$. The resulting equation becomes
\begin{equation}
\bigtriangledown _{9}^{2}Y=Q,\quad  \label{laplacian}
\end{equation}%
where $\bigtriangledown _{9}^{2}$ is the Laplacian on the 9-d flat space: $%
\bigtriangledown _{9}^{2}=[r_{1}^{-2}\partial _{1}(r_{1}^{2}\partial
_{1})+r_{2}^{-5}\partial _{2}(r_{2}^{5}\partial _{2})]$ and $Q=2\kappa
^{2}\tau _{D2}g_{s}^{-2}pr_{0}^{2}r_{1}^{-2}\delta ^{^{\prime
}}(r_{1}-r_{0})\delta ^{6}(r_{2})$ is the source term, where $\delta
^{^{\prime }}(r_{1}-r_{0})$\ is the derivative of $\delta (r_{1}-r_{0})$
with respect to $r_{1}$. $Y$ can be solved by integration via Green's
function:\
\begin{equation}
Y(\overrightarrow{r})=\ \int G(\overrightarrow{r},\overrightarrow{r}%
^{^{\prime }})Q(\overrightarrow{r}^{^{\prime }})d^{9}\overrightarrow{r}%
^{^{\prime }},  \label{integration-green}
\end{equation}%
where $G(\overrightarrow{r},\overrightarrow{r}^{^{\prime }})$\ is the green
function in the 9-d flat space defined as%
\begin{eqnarray}
\bigtriangledown _{9}^{2}G(\overrightarrow{r},\overrightarrow{r}^{^{\prime
}}) &=&\delta ^{9}(\overrightarrow{r}-\overrightarrow{r}^{^{\prime }}), \\
G(\overrightarrow{r},\overrightarrow{r}^{^{\prime }}) &=&\frac{C}{\left\vert
\overrightarrow{r}-\overrightarrow{r}^{^{\prime }}\right\vert ^{7}},
\end{eqnarray}%
where $C$\ is a constant.

Now let's study the $Y$\ at a point $(0,0,r_{1},0,0,0,0,0,r_{2})$. It is the
superposition of all the potentials generated by the sources at points
parameterized by ($r_{1}^{^{\prime }}\sin \theta \cos \varphi $, $%
r_{1}^{^{\prime }}\sin \theta \sin \varphi $, $r_{1}^{^{\prime }}\cos \theta
$, $0$,$0$,$0$,$0$,$0$,$0$), where the first three coordinates denote $%
x_{1},x_{2},x_{3}$ and the last six denote $x_{4},...,x_{9}$. From equation (%
\ref{integration-green}), now $Y$ should be:
\begin{eqnarray}
Y &=&\int \frac{C}{(r_{1}^{^{\prime }2}-2r_{1}r_{1}^{^{\prime }}\cos \theta
+r_{1}^{2}+r_{2}^{2})^{7/2}}2\kappa ^{2}\tau
_{D2}g_{s}^{-2}pr_{0}^{2}r_{1}^{^{\prime }-2}\delta ^{^{\prime
}}(r_{1}^{^{\prime }}-r_{0})\delta ^{6}(r_{2}^{^{\prime }})d^{6}%
\overrightarrow{r}_{2}^{^{\prime }}r_{1}^{^{\prime }2}dr_{1}^{^{\prime
}}\sin \theta d\theta d\varphi  \notag \\
&=&\int \frac{4\pi C\kappa ^{2}\tau _{D2}g_{s}^{-2}pr_{0}^{2}}{%
5r_{1}r_{1}^{^{\prime }}}\left( \frac{1}{[(r_{1}-r_{1}^{^{\prime
}})^{2}+r_{2}^{2}]^{5/2}}-\frac{1}{[(r_{1}+r_{1}^{^{\prime
}})^{2}+r_{2}^{2}]^{5/2}}\right) \delta ^{^{\prime }}(r_{1}^{^{\prime
}}-r_{0})dr_{1}^{^{\prime }},
\end{eqnarray}%
where we first integrated over $d^{6}\overrightarrow{r}_{2}^{^{\prime }}$, $%
d\theta $, $d\varphi $. Then we use $\delta ^{^{\prime }}(r_{1}^{^{\prime
}}-r_{0})=\underset{\epsilon \rightarrow 0}{\lim }\frac{1}{2\epsilon }%
[\delta (r_{1}^{^{\prime }}-r_{0}+\epsilon )-\delta (r_{1}^{^{\prime
}}-r_{0}-\epsilon )]$ and perform the integration and then take $\epsilon
\rightarrow 0$. \ We then get the result for $Y$ in equation (\ref{Y-result}%
) in section 4.

In the region where $r_{1}\gg r_{0},$\ we can expand $Y$ in terms of powers
of $r_{0}$:
\begin{equation}
Y=\frac{4\pi C\kappa ^{2}\tau _{D2}g_{s}^{-2}p}{5r_{1}}\left[ \frac{4}{3!}%
\left( \frac{105r_{1}}{r^{9}}-\frac{315r_{1}^{3}}{r^{11}}\right) r_{0}^{3}+%
\frac{8}{5!}\left( \frac{-4527r_{1}}{r^{11}}+\frac{34650r_{1}^{3}}{r^{13}}-%
\frac{45045r_{1}^{5}}{r^{15}}\right) r_{0}^{5}+O(r_{0}^{7})\right] .
\label{Y-expansion}
\end{equation}

We see the leading term of (\ref{Y-expansion}) is of order $r_{0}^{3}$: $Y=%
\frac{4\pi C\kappa ^{2}\tau _{D2}g_{s}^{-2}p}{5r_{1}}\left[ \frac{4}{3!}%
\left( \frac{105r_{1}}{r^{9}}-\frac{315r_{1}^{3}}{r^{11}}\right) r_{0}^{3}%
\right] =k\left( \frac{1}{r^{9}}-\frac{3r_{1}^{2}}{r^{11}}\right) r_{0}^{3},$
where $k$ is a constant. Then we get $h=k\frac{r_{1}^{3}r_{0}^{3}}{3r^{9}}$
and the leading terms of $H_{3}^{(II)}$ and $G_{6}^{(II)}$ are
\begin{eqnarray}
H_{3}^{(II)} &=&g_{s}\ast _{9}G_{6}^{(1)}=g_{s}k\frac{r_{0}^{3}}{r^{9}}\left[
\left( 1-3\frac{r_{1}^{2}}{r^{2}}\right) dr_{1}\wedge w_{2}-3\frac{r_{1}r_{2}%
}{r^{2}}dr_{2}\wedge w_{2}\right]  \notag \\
&=&g_{s}k\frac{r_{0}^{3}}{r^{9}}[T_{3}-3V_{3}], \\
G_{6}^{(II)} &=&k\frac{r_{0}^{3}}{r^{9}}[\ast _{9}T_{3}-\ast _{9}3V_{3}].
\end{eqnarray}%
We see that the leading terms of $H_{3}^{(II)}$ and $G_{6}^{(II)}$ in large $%
r$ region is precisely our $n=-9$\ solution (\ref{n=-9}) in section 2. Since
$r_{0}\propto \mu $, they are of the third order in $\mu $ in large $r$
region.

%------------------ References -------------------------


\begin{thebibliography}{99}
\bibitem{adscft} J. M. Maldacena, \textquotedblleft The large $N$ limit of
superconformal field theories and supergravity,\textquotedblright\ Adv.
Theor. Math. Phys. \textbf{2} (1998) 231, hep-th/9711200; S. S. Gubser, I.
R. Klebanov and A. M. Polyakov, \textquotedblleft Gauge theory correlators
from non-critical string theory,\textquotedblright\ Phys. Lett. B \textbf{428%
} (1998) 105, hep-th/9802109; E. Witten, \textquotedblleft Anti-de Sitter
space and holography,\textquotedblright\ Adv. Theor. Math. Phys. \textbf{2}
(1998) 253, hep-th/9802150; O. Aharony, S. S. Gubser, J. Maldacena, H.
Ooguri, Y. Oz, \textquotedblleft Large N Field Theories, String Theory and
Gravity," Phys.Rept. \textbf{323} (2000) 183-386, hep-th/9905111.

\bibitem{bmn021} D. Berenstein, J. M. Maldacena and H. Nastase,
\textquotedblleft Strings in flat space and pp waves from $\mathcal{N}=4$
Super Yang Mills,\textquotedblright\ JHEP \textbf{0204} (2002) 013,
hep-th/0202021.

\bibitem{n1star} C. Vafa and E. Witten, \textquotedblleft A Strong coupling
test of S duality,\textquotedblright\ \ Nucl. Phys. \textbf{B431}, 3 (1994),
hep-th/9408074; R. Donagi and E. Witten, \textquotedblleft Supersymmetric
Yang-Mills Theory And Integrable Systems,\textquotedblright\ Nucl. Phys.
\textbf{B460}, 299 (1996), hep-th/9510101.

\bibitem{ps} J. Polchinski, M. J. Strassler, \textquotedblleft The String
Dual of a Confining Four-Dimensional Gauge Theory," hep-th/0003136.

\bibitem{penrose} R. Penrose, \textquotedblleft Any space-time has a plane
wave limit,\textquotedblright\ in Differential Geometry and Gravity, Reidel,
Dordrecht 1976, pp. 271; J. Kowalski-Glikman, \textquotedblleft Vacuum
States In Supersymmetric Kaluza-Klein Theory,\textquotedblright\ Phys. Lett.
B \textbf{134}, 194 (1984); C. M. Hull, \textquotedblleft Exact PP-Wave
Solutions Of Eleven-Dimensional Supergravity," Phys. Lett. B\textbf{\ 139}
(1984) 39; R. Gueven, \textquotedblleft Plane Waves In Effective Field
Theories Of Superstrings,\textquotedblright\ Phys. Lett. B \textbf{191}, 275
(1987); M. Blau, J. Figueroa-O'Farrill, C. Hull and G. Papadopoulus,
\textquotedblleft Penrose limits and maximal
supersymmetry,\textquotedblright\ Class. Quant. Grav. \textbf{19} (2002)
L87, hep-th/0201081;

\bibitem{superposition} J. Figueroa-O'Farrill, G. Papadopoulos,
\textquotedblleft Homogeneous Fluxes, Branes and a Maximally Supersymmetric
Solution of M-theory,\textquotedblright\ JHEP \textbf{0108} (2001) 036,
hep-th/0105308.

\bibitem{supermembrane} B. de Wit, J. Hoppe, H. Nicolai , \textquotedblleft
On The Quantum Mechanics Of Supermembranes", Nucl. Phys. \textbf{B305}
(1988) 545.

\bibitem{BMN perturb} K. Dasgupta, M. M. Sheikh-Jabbari, M. Van Raamsdonk,
\textquotedblleft Matrix Perturbation Theory For M-Theory On a
PP-Wave,\textquotedblright\ JHEP \textbf{0205} (2002) 056, hep-th/0205185.

\bibitem{protect} K. Dasgupta, M. M. Sheikh-Jabbari, M. Van Raamsdonk,
\textquotedblleft Protected Multiplets of M-Theory on a Plane
Wave," JHEP \textbf{0209} (2002) 021, hep-th/0207050.

\bibitem{plefkaspectrum} N. Kim, J. Plefka , \textquotedblleft On the
Spectrum of PP-Wave Matrix Theory," Nucl. Phys. \textbf{B643} (2002) 31-48,
hep-th/0207034; N. Kim, J.-H. Park, \textquotedblleft Superalgebra for
M-theory on a pp-wave," Phys. Rev. \textbf{D66} (2002) 106007,
hep-th/0207061.

\bibitem{matrixtheory} T. Banks, W. Fischler, S. H. Shenker and L. Susskind,
\textquotedblleft M theory as a matrix model: A
conjecture,\textquotedblright\ Phys. Rev. \textbf{D55} (1997) 5112,
hep-th/9610043; L. Susskind, \textquotedblleft Another conjecture about
M(atrix) theory,\textquotedblright\ hep-th/9704080.

\bibitem{kabat-taylor} D. Kabat, W. Taylor, \textquotedblleft Spherical
membranes in Matrix theory," Adv. Theor. Math. Phys. \textbf{2} (1998)
181-206, hep-th/9711078.

\bibitem{transverse5} J. M. Maldacena, M. M. Sheikh-Jabbari, M. Van
Raamsdonk, \textquotedblleft Transverse Fivebranes in Matrix Theory," JHEP
\textbf{0301} (2003) 038, hep-th/0211139.

\bibitem{SYMS3} N. Kim, T. Klose, J. Plefka, \textquotedblleft Plane-wave
Matrix Theory from N=4 Super Yang-Mills on $R\times S^{3}$," Nucl. Phys.
\textbf{B671} (2003) 359, hep-th/0306054.

\bibitem{mye053} R. C. Myers, \textquotedblleft
Dielectric-Branes,\textquotedblright\ JHEP \textbf{9912} (1999) 022,
hep-th/9910053.

\bibitem{D0dual} N. Itzhaki, J. M. Maldacena, J. Sonnenschein, S.
Yankielowicz, \textquotedblleft Supergravity and The Large N Limit of
Theories With Sixteen Supercharges," Phys. Rev. \textbf{D58} (1998) 046004,
hep-th/9802042.

\bibitem{INVASION} J. McGreevy, L. Susskind and N. Toumbas,
\textquotedblleft Invasion of the giant gravitons from anti-de Sitter
space,\textquotedblright\ JHEP \textbf{0006} (2000) 008, hep-th/0003075.

\bibitem{determinant} V. Balasubramanian, M. Berkooz, A. Naqvi, M. J.
Strassler, \textquotedblleft Giant Gravitons in Conformal Field Theory,"
JHEP \textbf{0204} (2002) 034, hep-th/0107119.

\bibitem{CJR} S. Corley, A. Jevicki and S. Ramgoolam, \textquotedblleft
Exact correlators of giant gravitons from dual $\mathcal{N}=4$ SYM
theory,\textquotedblright\ Adv. Theor. Math. Phys. \textbf{5} (2002)
809-839, hep-th/0111222.

\bibitem{noncomu} M. Alishahiha, Y. Oz and M. M. Sheikh-Jabbari,
\textquotedblleft Supergravity and large N noncommutative field
theories,\textquotedblright\ JHEP \textbf{9911} (1999) 007, hep-th/9909215.

\bibitem{noncomutative} J. G. Russo, A. A. Tseytlin, \textquotedblleft
Waves, boosted branes and BPS states in M-theory," Nucl. Phys. \textbf{B490}
(1997) 121-144, hep-th/9611047; A. Hashimoto, N. Itzhaki, \textquotedblleft
Non-Commutative Yang-Mills and the AdS/CFT Correspondence," Phys. Lett. B
\textbf{465} (1999) 142-147, hep-th/9907166; J. M. Maldacena, J. G. Russo,
\textquotedblleft Large N Limit of Non-Commutative Gauge Theories," JHEP
\textbf{9909} (1999) 025, hep-th/9908134; R.~G.~Cai and N.~Ohta, ``Noncommutative and ordinary super Yang-Mills on (D(p-2),Dp) bound  states,'' JHEP {\bf 0003}, 009 (2000), hep-th/0001213.
%%CITATION = HEP-TH 0001213;%%


\bibitem{pbrane} G. Horowitz, A. Strominger, \textquotedblleft Black Strings
And P-Branes," Nucl. Phys. \textbf{B360} (1991) 197.

\bibitem{duff} M. J. Duff, J. X. Lu, \textquotedblleft Type II p-branes: the
brane-scan revisited," Nucl. Phys. \textbf{B390} (1993) 276-290,
hep-th/9207060.

\bibitem{boundarybulk} V. Balasubramanian, P. Kraus and A. Lawrence,
\textquotedblleft Bulk vs. boundary dynamics in anti-de Sitter
spacetime,\textquotedblright\ Phys. Rev. \textbf{D59}, 046003 (1999),
hep-th/9805171; T. Banks, M. R. Douglas, G. T. Horowitz and E. Martinec,
\textquotedblleft AdS dynamics from conformal field
theory,\textquotedblright\ hep-th/9808016.

\bibitem{superp} N. Iizuka, \textquotedblleft Supergravity, Supermembrane
and M(atrix) model on PP-Waves," Phys. Rev. \textbf{D68} (2003) 126002,
hep-th/0211138.

\bibitem{grana} M. Gra\~{n}a, J. Polchinski, \textquotedblleft
Supersymmetric Three-Form Flux Perturbations on $AdS_{5}$," Phys. Rev.
\textbf{D63} (2001) 026001, hep-th/0009211.

\bibitem{pssusynew} G. L. Cardoso, G. Curio, G. Dall'Agata, D. Lust,
\textquotedblleft Gaugino Condensation and Generation of Supersymmetric
3-Form Flux," hep-th/0406118.

\bibitem{susyrule} S. F. Hassan, \textquotedblleft T-Duality, Space-time
Spinors and R-R Fields in Curved Backgrounds," Nucl. Phys. \textbf{B568}
(2000) 145-161, hep-th/9907152; E. Bergshoeff, \textquotedblleft p-Branes,
D-Branes and M-Branes," hep-th/9611099; D. Marolf, L. Martucci, P. J. Silva,
\textquotedblleft Actions and Fermionic symmetries for D-branes in bosonic
backgrounds," JHEP \textbf{0307} (2003) 019, hep-th/0306066.

\bibitem{g=1} M. J. Duff, J. T. Liu, J. Rahmfeld, \textquotedblleft g=1 for
Dirichlet 0-branes," Nucl. Phys. \textbf{B524 }(1998) 129-140,
hep-th/9801072.

\bibitem{d0multipole} K. Millar, W. Taylor, M. Van Raamsdonk,
\textquotedblleft D-particle polarizations with multipole moments of
higher-dimensional branes," hep-th/0007157.

\bibitem{puzzle} I. Bena, D. Smith, \textquotedblleft Towards the solution
to the giant graviton puzzle," hep-th/0401173.

\bibitem{vafa}
C.~Vafa, ``Superstrings and topological strings at large N,'' J.\ Math.\ Phys.\  {\bf 42}, 2798 (2001), hep-th/0008142; I.~R.~Klebanov and M.~J.~Strassler, ``Supergravity and a confining gauge theory: Duality cascades and chiSB-resolution of naked singularities,'' JHEP {\bf 0008}, 052 (2000), hep-th/0007191.
%%CITATION = HEP-TH 0008142;%%
%%CITATION = HEP-TH 0007191;%%

\bibitem{othervacua} K. Sugiyama, K. Yoshida, \textquotedblleft
Supermembrane on the PP-wave Background," Nucl. Phys. \textbf{B644} (2002)
113-127, hep-th/0206070; K. Sugiyama, K. Yoshida, \textquotedblleft BPS
Conditions of Supermembrane on the PP-wave," Phys. Lett. B \textbf{546}
(2002) 143-152, hep-th/0206132; G. Bonelli, \textquotedblleft Matrix Strings
in pp-wave backgrounds from deformed Super Yang-Mills Theory," JHEP \textbf{%
0208} (2002) 022, hep-th/0205213; N. Kim , K. M. Lee, P. Yi,
\textquotedblleft Deformed Matrix Theories with N=8 and Fivebranes in the PP
Wave Background," JHEP \textbf{0211} (2002) 009, hep-th/0207264; K.
Sugiyama, K. Yoshida, \textquotedblleft Giant Graviton and Quantum Stability
in Matrix Model on PP-wave Background," Phys. Rev. \textbf{D66} (2002)
085022, hep-th/0207190; A. Mikhailov, \textquotedblleft Nonspherical Giant
Gravitons and Matrix Theory," hep-th/0208077; J.-H. Park, \textquotedblleft
Supersymmetric objects in the M-theory on a pp-wave," JHEP \textbf{0210}
(2002) 032, hep-th/0208161.

\bibitem{bak033} D. Bak, \textquotedblleft Supersymmetric Branes in PP-Wave
Background,\textquotedblright\ Phys. Rev. \textbf{D67} (2003) 045017,
hep-th/0204033.

\bibitem{shin} S. Hyun, H. Shin, \textquotedblleft Branes from Matrix Theory
in PP-Wave Background," Phys. Lett. B \textbf{543} (2002) 115-120,
hep-th/0206090.

\bibitem{jeremy1}
S.~R.~Das, J.~Michelson and A.~D.~Shapere, ``Fuzzy spheres in pp-wave matrix string theory,''
Phys.\ Rev.\ D {\bf 70}, 026004 (2004), hep-th/0306270.
%%CITATION = HEP-TH 0306270;%%

\bibitem{instanton} J.-T. Yee , P. Yi, \textquotedblleft Instantons of
M(atrix) Theory in PP-Wave Background," JHEP \textbf{0302} (2003) 040,
hep-th/0301120.

\bibitem{nahm} C. Bachas, J. Hoppe, B. Pioline, \textquotedblleft Nahm's
equations, $\mathcal{N}=1^{\ast }$ domain walls, and D-strings in $%
AdS_{5}\times S_{5}$," JHEP \textbf{0107} (2001) 041, hep-th/0007067; A.
Frey, \textquotedblleft Brane Configurations of BPS Domain Walls for the $%
\mathcal{N}=1^{\ast }$ SU(N) Gauge Theory," JHEP \textbf{0012} (2000) 020,
hep-th/0007125.

\bibitem{benaetal} I. Bena, A. Nudelman, \textquotedblleft Warping and vacua
of (S)YM$_{2+1}$," Phys. Rev. \textbf{D62} (2000) 086008, hep-th/0005163; I.
Bena, A. Nudelman, \textquotedblleft Exotic polarizations of D2 branes and
oblique vacua of (S)YM$_{2+1}$," Phys. Rev. \textbf{D62} (2000) 126007,
hep-th/0006102; I. Bena, \textquotedblleft The M-theory dual of a 3
dimensional theory with reduced supersymmetry," Phys. Rev. \textbf{D62}
(2000) 126006, hep-th/0004142; I. Bena, C. Ciocarlie, \textquotedblleft
Exact N=2 Supergravity Solutions With Polarized Branes," hep-th/0212252; I.
Bena, D. Vaman, \textquotedblleft The polarization of M5 branes and little
string theories with reduced supersymmetry," JHEP \textbf{0111} (2001) 032,
hep-th/0101064.

\bibitem{otherbrane} F. Zamora, \textquotedblleft Non-Supersymmetric
SO(3)-Invariant Deformations of $\mathcal{N}=1^{\ast }$ Vacua and their Dual
String Theory Description," JHEP \textbf{0012} (2000) 021, hep-th/0007082;
C. Ahn, T. Itoh, \textquotedblleft Dielectric-branes in Non-supersymmetric
SO(3)-invariant Perturbation of Three-dimensional N=8 Yang-Mills Theory,"
Phys. Rev. \textbf{D64} (2001) 086006, hep-th/0105044.

\bibitem{dieletric} I. Bena, N. P. Warner, \textquotedblleft A harmonic
family of dielectric flow solutions with maximal supersymmetry,"
hep-th/0406145.

\bibitem{inflation} O. DeWolfe, S. Kachru, H. Verlinde, \textquotedblleft
The Giant Inflaton," JHEP \textbf{0405} (2004) 017, hep-th/0403123.

\bibitem{general-ppwave} M. Cvetic, H. Lu, C.N. Pope, \textquotedblleft
M-theory PP-waves, Penrose Limits and Supernumerary Supersymmetries," Nucl.
Phys. \textbf{B644} (2002) 65-84, hep-th/0203229; J. P. Gauntlett , C. M.
Hull, \textquotedblleft pp-waves in 11-dimensions with extra supersymmetry,"
JHEP \textbf{0206} (2002) 013, hep-th/0203255; J. Michelson, "A pp-Wave With
26 Supercharges," Class. Quant. Grav. \textbf{19} (2002) 5935-5949,
hep-th/0206204; K. M. Lee, \textquotedblleft M-theory on Less Supersymmetric
PP-Waves," Phys. Lett. B \textbf{549} (2002) 213-220, hep-th/0209009; N.
Ohta , M. Sakaguchi, \textquotedblleft Uniqueness of M-theory PP-Wave
Background with Extra Supersymmetries," Phys. Rev. \textbf{D69} (2004)
066006, hep-th/0305215.

\bibitem{jeremy2}
J.~Michelson, ``Matrix theory of pp waves,'' hep-th/0401050.
%%CITATION = HEP-TH 0401050;%%

\bibitem{iia} L. J. Romans, \textquotedblleft Massive N=2a Supergravity In
Ten-Dimensions," Phys. Lett. B \textbf{169} (1986) 374; C. P. Herzog, I. R.
Klebanov, \textquotedblleft Gravity Duals of Fractional Branes in Various
Dimensions," Phys. Rev. \textbf{D63} (2001) 126005, hep-th/0101020, Appendix
A.

\bibitem{addsource} P. Rajan, \textquotedblleft D2-brane RR-charge on
SU(2)," Phys. Lett. B \textbf{533} (2002) 307-312, hep-th/0111245.
\end{thebibliography}
\end{document}